\documentclass[12pt]{article}
\usepackage{epsfig,amsfonts,amssymb}
\usepackage{hyperref}
\usepackage{cite}
\input epsf.sty
\usepackage{cite}

\def\ZZZ{{\hbox{ Z\kern-1.6mm Z}}}
\def\RRR{{\hbox{ R\kern-2.4mm R}}}
\def\CCC{{\hbox{ C\kern-2.0mm C}}}
\def\zzz{{\hbox{z\kern-1mm z}}}

\newcommand{\ten}{{(10)}}

\newcommand{\nn}{\nonumber \\}

\newcommand{\qeq}{{\hbox{=\kern-2.3mm ? \kern.5mm }}}
\renewcommand{\qeq}{=}

\newcommand{\eps}{\epsilon}
\newcommand{\vareps}{\varepsilon}

\newcommand{\vp}{\varphi}
\newcommand{\ve}{\varepsilon}

\newcommand{\DD}{{\cal D}}

\newcommand{\AAA}{{\cal A}}

\newcommand{\FF}{{\cal F}}

\newcommand{\OO}{{\cal O}}

\newcommand{\LL}{{\cal L}}

\newcommand{\wt}{\widetilde}
\newcommand{\wh}{\widehat}

\newcommand{\NN}{{\cal N}}

\newcommand{\SSS}{{\cal S}}

\newcommand{\tk}{\tilde \kappa}

\newcommand{\be}{\begin{equation}}
\newcommand{\ee}{\end{equation}}
\newcommand{\ben}{\begin{eqnarray}\displaystyle}
\newcommand{\een}{\end{eqnarray}}

\newcommand{\refb}[1]{(\ref{#1})}
\newcommand{\p}{\partial}
\newcommand{\sectiono}[1]{\section{#1}\setcounter{equation}{0}}
\newcommand{\subsectiono}[1]{\subsection{#1}\setcounter{equation}{0}}

\def\one{{\hbox{ 1\kern-.8mm l}}}
\def\zero{{\hbox{ 0\kern-1.5mm 0}}}

\renewcommand{\theequation}{\thesection.\arabic{equation}}

\begin{document}

\baselineskip 24pt

\begin{center}
{\Large \bf  Logarithmic Corrections to Extremal
Black Hole Entropy
from Quantum Entropy Function}

\end{center}

\vskip .6cm
\medskip

\vspace*{4.0ex}

\baselineskip=18pt

\centerline{\large \rm    Shamik Banerjee$^a$, Rajesh K. 
Gupta$^a$ and Ashoke
Sen$^{a,b}$}

\vspace*{4.0ex}

\centerline{\large \it $^a$Harish-Chandra Research Institute}
\centerline{\large \it  Chhatnag Road, Jhusi,
Allahabad 211019, India}
\centerline{and}
\centerline{\large \it 
$^b$LPTHE, Universite Pierre et Marie Curie, Paris 6}
\centerline{\large \it 
4 Place Jussieu,  75252 Paris Cedex 05, France}

\vspace*{1.0ex}
\centerline{E-mail:  bshamik@mri.ernet.in, rajesh@mri.ernet.in,
sen@mri.ernet.in}

\vspace*{5.0ex}

\renewcommand{\check}{\bar }

\centerline{\bf Abstract} \bigskip

We evaluate the one loop determinant of matter 
multiplet fields of
$\NN=4$ supergravity 
in the near horizon geometry of quarter BPS black holes, and
use it to calculate 
logarithmic corrections to the entropy of these black holes
using the quantum entropy function formalism.
We show that even though individual fields give non-vanishing
logarithmic contribution to the entropy, the net contribution from
all the fields in the matter multiplet vanishes. 
Thus logarithmic corrections to the entropy of quarter BPS
black holes, if present, must be independent of the number of
matter multiplet fields in the theory.
This is consistent with
the microscopic results. 
During our analysis we also determine the complete spectrum of small
fluctuations of matter multiplet fields in the near horizon geometry.

\vfill \eject

\baselineskip=18pt

\tableofcontents

\sectiono{Introduction} \label{s1}

Wald's formula gives a generalization of the 
Bekenstein-Hawking entropy
in a classical theory of gravity with higher derivative 
terms, possibly coupled to other matter 
fields\cite{9307038,9312023,9403028,9502009}.
In the extremal limit this leads to a simple algebraic 
procedure for determining the near horizon field
configurations and the entropy\cite{0506177,0508042}, 
leading to a simple
proof of the attractor 
mechanism\cite{9508072,9602111,9602136} in a general
higher derivative theory coupled to matter. 

Given this success, one could ask: is there a generalization of the
Wald's formula to the full quantum theory? 
At least for extremal BPS
black holes, there is reason to expect that such a formula might exist,
since on the microscopic side there is a precise result for the
degeneracy (more precisely an appropriate 
index\footnote{See ref.\cite{0903.1477,appear} for a 
discussion on how the
black hole entropy can be related to an index.
In this paper we shall not distinguish between index
and degeneracy.}) for these BPS
black holes.
One naive approach to this problem will be to continue to use 
Wald's formula by replacing the classical action by the
one particle irreducible (1PI) action. In string theory
this approach has been successful
in a number of cases, producing highly non-trivial dependence of the
entropy on the charges which can then be verified by explicit computation
of the statistical entropy in a microscopic 
description\cite{0412287,0510147,0605210,0609109}.
(Further results on the microscopic spectrum of $\NN=4$
supersymmetric string theories, which will be our focus
of attention, can be found 
in
\cite{9607026,0505094,0506249,0508174,
0602254,0603066,0607155,0612011,
0702141,0702150,0705.1433,0705.3874,0706.2363,
0708.1270,0802.0544,0802.1556,0803.2692,0806.2337,
0807.4451,0809.4258,0808.1746,0901.1758,0907.1410,
0911.0586,0911.1563,1002.3857,suresh}.
Early studies on the 
macroscopic entropy of these black holes 
can be found
in \cite{9507090,9512031,9711053,9812082,9906094,
0007195}).

There is however a simple reason why this cannot be the
complete prescription. In its original formulation,
Wald's formula holds only for local action. On the other hand the
1PI action at sufficiently high orders in derivatives contains non-local
terms due to the presence of the massless fields in the supergravity
theory. 
Thus the prescription of replacing the classical action
by the 1PI action in Wald's formula
cannot be the complete story.\footnote{See \cite{1003.1083} 
for an
attempt to resolve this using an auxiliary scalar field.}
A proposal for overcoming this  difficulty based on a
Euclidean path integral approach was suggested in
\cite{0809.3304}. In this formulation, 
called the quantum entropy function
formalism,
the degeneracy associated with the black hole
horizon is given by the finite part of the string theory
partition function
in the near horizon geometry of the black hole containing an
$AdS_2$ factor.  
More precisely, the partition function is calculated by evaluating
the string theory path integral over all string field
configurations subject to
the condition that near the asymptotic boundary of $AdS_2$ the
configuration approaches 
the near horizon geometry of the extremal black hole
under consideration. Such a partition function is divergent
due to the infinite volume of $AdS_2$, but the rules of 
$AdS_2/CFT_1$ correspondence gives a precise procedure
for removing this divergence.\footnote{Technically this is
identical to the procedure one follows for removing the quark
self-energy divergence while computing the Wilson /'t Hooft 
line expectation
values in gauge theories via holographic 
method\cite{9803002,9803001,0904.4486}, 
but whether
there is a deeper physical connection between these two
quantities remains to be seen.}
While in the classical limit this prescription gives us
back the exponential of the Wald action, it can in principle be
used to systematically calculate the quantum corrections to the
entropy of an extremal black hole. Indeed many of the
non-perturbative features of the known 
spectrum of quarter BPS states in $\NN=4$
supersymmetric string theories have been reproduced 
from the macroscopic side using this
formalism\cite{0810.3472,0903.1477,0904.4253,
0908.0039,0911.1563,1002.3857}. 
These non-perturvative effects arise from
inclusion in the path integral the contribution from
non-trivial saddle points which have the same asymptotic geometry
as the near horizon geometry of the black hole, but differ
from it in the interior of $AdS_2$.

In order to make full use of this program we need to carry out the
path integral over the string fields around each saddle point. We can take
two approaches to this problem. The simplicity of the microscopic
formula for the black hole entropy in $\NN=4$ and $\NN=8$
supersymmetric string theories leads us to
expect that the contribution to the path integral
from each saddle point
can be expressed as a finite dimensional integral with simple
integrand. Given the large amount of supersymmetry possessed
by the near horizon geometry, one could try to achieve this
using localization techniques\cite{heckman,wittens,witten92,
9112056,9204083,9511112,zaboronsky,0206161,
0712.2824,ati1,ati2,0608021}. 
In particular it is quite conceivable that
supersymmetry will help us  localize the path integral 
over string fields 
to a finite dimensional subspace of the full configuration space,
which could then be directly compared with the corresponding
contribution to the microscopic formula\cite{0905.2686}.
On the other hand one could also take a brute
force approach and try to evaluate the path integral over string fields
in perturbation theory around each saddle point.
This can then be compared with a similar expansion of the
microscopic degeneracy formula in appropriate inverse powers of
the charges.

The analysis of this paper will be based
on the second approach. We shall 
calculate the one loop contribution to the
quantum entropy function to analyze one
specific feature of the entropy formula, --
logarithmic corrections to the classical entropy.
More precisely, we shall consider the limit in which
all charges become uniformly large, carrying a common scale
$\Lambda$, and study corrections of order $\ln\Lambda$ to the
entropy.\footnote{This is to be distinguished from the Cardy limit
in which one of the charges representing momentum along an
internal circle becomes large. In this limit the logarithmic correction
to the black hole entropy is known to be 
universal\cite{0002040,0005017,0104010}.} 
The motivation for this study comes from the
known results on the microscopic spectrum of the quarter
BPS dyons in $\NN=4$ supersymmetric string theories and
1/8 BPS dyons in $\NN=8$ supersymmetric string theories.
If we denote by $\Delta$ the unique quartic combination of the
charges which is invariant under continuous U-duality group
of these theories, then for large $\Delta$ the microscopic entropy
$S_{micro}$, computed by taking the logarithm of the
appropriate helicity trace index\cite{9708062,9708130}, grows as
\be \label{esmicro}
S_{micro} = \cases{\pi \sqrt{\Delta} + \OO(1) \quad \hbox{for}
\quad \NN=4\cr
\pi \sqrt{\Delta} - 2 \ln\Delta + \OO(1) \quad \hbox{for}
\quad \NN=8}\, .
\ee
The result for the $\NN=4$ theory can be found 
in \cite{0412287,0510147,0609109} and that for $\NN=8$ theory
can be found in \cite{0908.0039}. Thus in the limit
described above the quarter BPS dyons in $\NN=4$ supersymmetric
theories have no logarithmic corrections whereas 1/8 BPS dyons in
$\NN=8$ supersymmetrc theories have corrections of order
$-8\ln\Lambda$.
Our goal will be to understand some aspects of 
these results from
the macroscopic viewpoint.

We shall begin by trying to understand the origin of possible
logarithmic corrections to the entropy in the quantum entropy
function formalism.
As we shall see, for this study the contribution from the
stringy modes -- and the Kaluza-Klein modes associated
with the internal directions --
become irrelevant, and we only need to compute the
contribution from the massless modes living on the near
horizon $AdS_2\times S^2$ geometry. 
As a simple exercise we first
calculate the one loop determinant of a
massless scalar field in the near horizon 
$AdS_2\times S^2$ background using heat kernel
method\footnote{Similar calculations in $AdS_3$ background,
with a somewhat different application in mind, can be found
in \cite{0804.1773,0911.5085}.}, and show that after following
the prescription of extracting the entropy from the one loop
partition function, we do generate a logarithmic correction to the
entropy. Furthermore this agrees with earlier result of
\cite{9709064} calculated using a somewhat different approach (more
detailed discussion on the comparison with other approaches
will be given below). 

Applying this procedure to compute logarithmic corrections to
string theoretic black holes requires us to evaluate the one
loop contribution to the partition function from the fluctuation
of massless fields in the attractor geometry. The main
technical difficulty in this computation is the diagonalization
of the kinetic terms of various fields. Since the background contains
electric and magnetic fields besides gravity, and since the supergravity
action is non-linear, the fluctuations of scalars, vectors and metric
(and similarly of spin 1/2 and spin 3/2 fields) mix with each other.
However for quarter BPS black holes in $\NN=4$ supersymmetric
string theories, which will be the main focus of our
analysis, there is a simplification: the near horizon background
involves purly gravitational and graviphoton fields, but no matter
multiplet fields. Due to this property the quadratic terms in the
action expanded around this background do not contain any
mixing term between matter and gravity multiplet fields. This
allows us to analyze the contribution to the partition
function from the matter multiplet fields and gravity multiplet
fields separately.

In this paper we study the contribution to the partition
function due to the fluctuations
of the matter multiplets. The first step in 
this process is to find explicitly all
the eigenvalues and eigenfunctions of the kinetic operator
acting on the matter multiplet, both in the bosonic and the
fermionic sectors. We then express the
one loop contribution to the partition function
in terms of this data, and find that in the final expression the
contribution from the bosonic and the fermionic fields cancel.
While a similar calculation is possible in principle for the
fields in the gravitational sector, the computation is technically
involved, and we have not carried out this analysis. 
As a result for any single theory we cannot make a definite
macroscopic prediction for the black hole entropy. However if
we consider a collection of different $\NN=4$ supersymmetric
string theories then our result has a definite prediction,
namely that the logarithmic correction
to the black hole entropy in $\NN=4$ supersymmetric string
theories is independent of the number of matter multiplets we
have in the theory. 
This is borne out in the microscopic analysis,
-- the net logarithmic correction being zero
in all known $\NN=4$ supersymmetric string 
theories irrespective of the number of matter
multiplets the theory
contains\cite{0412287,0510147,0605210,0609109}. 
In order to fully reproduce the results given
in \refb{esmicro} we shall have to compute the
contribution from the gravity multiplet in $\NN=4$
and $\NN=8$ supersymmetric
string theories.

To put our results in context we note that part of the one loop
contribution to the entropy of BPS black holes has been
analyzed earlier, leading to non-trivial agreement between the
microscopic and the macroscopic 
results\cite{0412287,0510147,0609109}.
These results were computed using the local part of the
one loop effective action derived in
\cite{9610237,9708062} for which one could use 
Wald's formula\cite{9307038} or equivalently the classical
entropy function formalism\cite{0506177}. This local effective
action, computed in type IIB string theory on $K3\times T^2$
and its various orbifolds, included contribution from massive
string states carrying winding and momentum along the
cycles of $T^2$, but the
contribution due to the massless modes had to be removed
by hand so as to avoid infrared divergent results.
In contrast our analysis in this paper computes
part of the contribution from the massless sector. Thus this
contribution must be added to the result of the 
previous analysis.

Logarithmic corrections to the (extremal) 
black hole entropy have been analyzed before from different points
of 
view\cite{9412161,9604118,9709064,0002040,0005017,
0104010,0406044,0409024,0805.2220,
0808.3688,0809.1508,0911.4379,1003.1083}.
The previous approaches can be divided into two
broad classes, -- 
microscopic and 
macroscopic. In the
microscopic analysis 
the logarithmic corrections are computed
by using specific microscopic description of the theory,
while in the macroscopic approach the logartithmic corrections
are computed from the analysis of fluctuating quantum
fields in
a black hole background. 
The macroscopic approaches can also be
divided into two categories. In one category, that involves
entropy of BTZ black holes, one first analyzes the gravity path
integral in asymptotically $AdS_3$ spaces to compute the
partition function, and then computes the entropy by taking a
laplace transform of the partition function. In this approach
the logarithmic terms arise in the process of taking the Laplace
transform. In the second category
one computes the entropy directly
by analyzing the quantum fluctuations of various fields in the
black hole geometry. The analysis of this paper clearly
falls in the last category.

To be more specific, we shall now compare our method to
\cite{9709064} which is closest in spirit. In
\cite{9709064} logarithmic corrections to 
the entropy of extremal black hole was calculated by relating it
to the partition function of the theory in the near horizon geometry
with a conical defect. 
This requires computing the heat kernel
of various fields in a background with conical defect. 
However while attempting to make this 
into a general prescription for
computing black hole entropy in string theory, 
one runs into the problem that
string theory may not
make sense in backgrounds with
arbitrary conical defects other than those obtained by taking 
orbifolds of smooth space-time.
Our
approach also requires computing the heat kernel of various
fields, but directly in the near horizon geometry without a
conical defect. 
As a result it is completely well defined once we adopt the
infrared subtraction procedure described in 
\cite{0809.3304}.  
Nevertheless
the results of our approach agree with those of \cite{9709064}
for cases where both methods have been applied {\it e.g.} for
a massles scalar field.
The main advantage of
our approach is that we begin with a general prescription for
computing the entropy of an extremal black hole based on
$AdS_2/CFT_1$ correspondence, and then evaluate it using
various approximations. 
This allows us to carry out a systematic comparison between
the macroscopic and microscopic entropies.
So far (including the results of this paper)
this comparison includes
classical Wald entropy and some one loop and non-perturbative results.

The rest of the paper is organised as follows. In
\S\ref{s2} we show how quantum entropy function can be used
to calculate logarithmic correction to the entropy of an extremal 
black hole due to a single massless scalar field coupled to the
background metric by minimal coupling. 
This requires computing the eigenvalues and eigenfunctions of
the scalar Laplacian in the near horizon geometry, and the heat
kernel constructed from these data.
We also study the
effect of introducing a mass term for the scalar, and show that
massive stringy states do not give any logarithmic correction
to the entropy of an extremal black hole. In \S\ref{sother} we generalize
the analysis to include contribution from massless vector,
$p$-form
and spinor fields coupled to the background metric via minimal
coupling. In \S\ref{sflux} we focus on the near horizon geometry of
quarter BPS black holes in $\NN=4$ supersymmetric string
theories. By expanding the supergravity action in this near
horizon background we find the complete quadratic action involving
the various fluctuating fields in the matter and gravity multiplet.
This action contains the minimal coupling of various fields
to the background metric, but also contains additional terms including
mixing between fields of different spin.
We find  however
that at the quadratic order there is no mixing between
the fluctuations in the matter and gravity multiplet fields, and
hence we can separately analyze the one loop contribution from the
two sets of fields. In \S\ref{sfive} we find the eigenvalues of the kinetic
operator in the matter multiplet (which in general contains
a mixing between the scalar and the vector fields, and also
between different components of the spin half field) and use this
to compute the one loop contribution to the quantum entropy
function. We find that while individual fields give logarithmic
contribution to the entropy, the total logarithmic contribution from
each vector multiplet vanishes, in agreement with the
microscopic results. We end in \S\ref{sdis} by summarizing the results
and speculating on the possible application of the pure spinor
formalism for a one loop computation of the quantum entropy function
in full string theory.
In appendix \ref{sa} we analyze the contribution from
integration over the zero modes which were left out from
the functional integral in the analysis of \S\ref{sfive} and show that
they do not give any additional 
logarithmic correction to the black hole
entropy.

\sectiono{Logarithmic correction
to the black hole entropy due to a single scalar field} \label{s2}

Suppose we have an extremal black hole with near horizon geometry
$AdS_2\times S^2$, with equal size $a$ of $AdS_2$ and $S^2$. Then
the Euclidean near horizon metric takes the form
\be \label{e1}
a^2 \left( d\eta^2 + \sinh^2\eta \, d\theta^2\right)
+ a^2 (d\psi^2 + \sin^2\psi \, d\phi^2) \, ,
\ee
where $\theta$ and $\phi$ are periodic coordinates with period
$2\pi$. We choose
the sign convention for the euclidean action $\SSS$
such that the weight
factor inserted into the path integral is given by $e^\SSS$.
Let $\Delta\LL_{eff}$ denote the one loop correction to the four
dimensional effective lagrangian density evaluated in the background
geometry \refb{e1}. Then the one loop correction to the action is
given by
\be \label{e2}
\Delta \SSS =
 \int\, \sqrt{\det g} \, d\eta \, d\theta  \, d\psi\,
 d\phi \, \Delta\LL_{eff}
= 8\pi^2 \, a^4 \, (\cosh\eta_0-1) \, \Delta\LL_{eff}
\ee
where $\eta_0$ is an infrared cut-off.
The term proportional to
$\cosh\eta_0$ has the interpretation of $-\beta \Delta E_0
+\OO\left(\beta^{-1}\right)$
where $\beta=2\pi a \sinh\eta_0$ is the inverse temperature given
by the length of the boundary of $AdS_2$
parametrized by $\theta$ and $\Delta E_0$
is the shift in the
ground state energy\cite{0809.3304,0903.1477}.
The rest of the contribution can be interpreted as the
one loop correction to the black hole 
entropy\cite{0809.3304,0903.1477} and takes the
form
\be \label{e3}
\Delta S_{BH} = -8\pi^2 a^4\, \Delta\, \LL_{eff}\, .
\ee
We shall now describe the general procedure for calculating
$\Delta\LL_{eff}$.

Let us assume that the theory contains a massless scalar field.
If we denote the eigenvalues of the scalar laplacian by $\{-\kappa_n\}$
and the corresponding normalized
eigenfunctions by $f_n(x)$ then the heat
kernel $K^s(x,x';s)$ of the scalar Laplacian is defined as
(see \cite{gilkey,0306138} and references therein)
\be \label{eh1}
K^s(x,x';s) = \sum_n \, e^{-\kappa_n\, s} \, f_n(x)\,
f_n(x')\, .
\ee
The superscript $s$ on $K$ reflects that the laplacian acts on
the scalar fields.
In \refb{eh1} we have assumed that we are working in a basis
in which the eigenfunctions are real; if this is not the case then we
need to take the complex conjugate of $f_n(x')$.
$K^s(x,x'; s)$  satisfies the equation
\be \label{e3a}
(\p_s - \square_x) K^s(x,x';s) = 0; \quad K^s(x,x'; s=0) 
= \delta^{(4)}(x-x')
\, ,
\ee
$\square_x$ being the Laplacian on $AdS_2\times S^2$.
The contribution of this scalar field to the one loop effective
action can now be expressed as
\be \label{e4}
\Delta\SSS =-{1\over 2}\, \sum_n \ln\kappa_n =
{1\over 2} \int_\eps^\infty {ds\over s} \sum_n e^{-\kappa_n s}
=
{1\over 2}\, \int_{\eps}^\infty\, {ds\over s} \, 
\int d^4 x \, \sqrt{\det g}\,  K^s(x,x; s)\, ,
\ee
where $g_{\mu\nu}$ is the $AdS_2\times S^2$ metric 
and $\eps$
is an ultraviolet cut-off.
Comparing this with \refb{e2} we see 
that\footnote{There are various other methods for evaluating functional
determinants in non-trivial space-time backgrounds, see
{\it e.g.} \cite{0305010, 0602106,0908.2657}. However the form
of the result given in \refb{e3ab} is closest to the form in which
we expect to obtain the answer in string theory, with the integration
over $s$ replaced by integration over the modular parameter of
a torus and $K^s(0;s)$ replaced by the torus partition function
of the first quantized string.}
\be \label{e3ab}
\Delta\LL_{eff} = {1\over 2} \, \int_{\eps}^\infty\, {ds\over s} \, 
K^s(0;s)\, ,
\ee
where $K^s(0;s)\equiv K^s(x, x; s)$. Note that
using the fact that $AdS_2$ and $S^2$ are homogeneous spaces
we have dropped the dependence on $x$ from $K^s(x,x;s)$.

Now it follows from
\refb{eh1} and the fact that
$\square_{AdS_2\times S^2}=\square_{AdS_2}+\square_{S^2}$
that the heat kernel of a massless scalar field
on $AdS_2\times S^2$ 
is given by the product
of the heat kernels on $AdS_2$ and $S^2$, and in the
$x'\to x$ limit takes the form\cite{campo}
\be \label{e4a}
 K^s(0;s) = K^s_{AdS_2}(0;s) K^s_{S^2}(0;s)\, .
\ee
$K^s_{S^2}$ and $K^s_{AdS_2}$ in turn can be calculated using
\refb{eh1} if we know the eigenfunctions and the eigenvalues of the
Laplace operator on these respective spaces. 
Fortunately these have been studied 
extensively\cite{campo,camhig1,campo2,camhig2}.
On $S^2$ the normalized
eigenfunctions of $-\square$ are just the usual spherical harmonics
$Y_{lm}(\psi,\phi)/a$ with eigenvalues $l(l+1)/a^2$. Since $Y_{lm}$
vanishes at $\psi=0$ for $m\ne 0$, and $Y_{l0}=\sqrt{2l+1}/\sqrt{4\pi}$ 
at $\psi=0$ we have
\be \label{e6p}
K^s_{S^2}(0;s) = {1\over 4\pi a^2} 
\sum_{l} e^{-sl(l+1)/a^2} (2l+1)\, .
\ee
On the other hand on $AdS_2$ the $\delta$-function
normalized eigenfunctions
of $-\square$ are given by\cite{camhig1}
\ben \label{e5p}
f_{\lambda,k}(\eta,\theta)
&=& {1\over \sqrt{2\pi\, a^2}}\, {1\over 2^{|k|} (|k|)!}\, \left|
{\Gamma\left(i\lambda +{1\over 2} + |k|\right)\over
\Gamma(i\lambda)}\right|\, 
e^{ik\theta} \sinh^{|k|}\eta\nn
&& F\left(i\lambda +{1\over 2}+|k|, -i\lambda 
+{1\over 2}+|k|; |k|+{1}; -\sinh^2{\eta\over 2}\right), \nn
&& \qquad \qquad \qquad \qquad
k\in \ZZZ, \qquad 0<\lambda<\infty\, ,
\een
with eigenvalue $\left({1\over 4}+\lambda^2\right)/a^2$. 
Here $F$ denotes hypergeometric function.
Since
the eigenfunction described in \refb{e5p} vanishes at
$\eta=0$ for $k\ne 0$, only the $k=0$ states will contribute to
$K^s_{AdS_2}(0;s)$. At $\eta=0$ the $k=0$ state
has the value
$\sqrt{\lambda\tanh(\pi\lambda)}/\sqrt{2\pi a^2}$.
Thus \refb{eh1} gives
\be \label{e5q}
K^s_{AdS_2}(0;s) =  {1\over 2\pi\, a^2}
\int_0^\infty \, d\lambda \, \lambda
\tanh(\pi\lambda) \, 
\exp\left[- s\left(\lambda^2 +{1\over 4}\right)/a^2\right]\, 
\, .
\ee
Combining \refb{e6p} and \refb{e5q} we get the heat kernel of a
scalar field on $AdS_2\times S^2$:
\be \label{ecomb1}
K^s(0;s) 
= {1\over 8\pi^2 a^4}\,
\sum_{\ell=0}^\infty (2l+1) \int_0^\infty \, d\lambda \, \lambda
\tanh(\pi\lambda) \, \exp\left[ -\bar s \lambda^2 - \bar 
s \left(l+{1\over 2}
\right)^2
\right]\, ,
\ee
where
\be \label{e7}
\bar s = s/a^2\, .
\ee
The associated eigenstates of the laplacian operator on $AdS_2\times S^2$
are obtained by taking the products of the spherical harmonics and the
function $f_{\lambda,k}$ given in \refb{e5p} and satisfy
\be \label{eigen1}
\square\, f_{\lambda,k}(\eta,\theta) \, Y_{lm}(\psi, \phi)
= -{1\over a^2}\, 
\left\{l(l+1) + \lambda^2 +{1\over 4} \right\}
\, f_{\lambda,k}(\eta,\theta) \, Y_{lm}(\psi, \phi)\, .
\ee

We can in principle evaluate the full one loop effective action 
due to massless fields using
\refb{e3ab} and \refb{ecomb1}, 
but our goal is to extract the piece proportional
to $\ln a$ for large $a$. 
Such contributions come from the region of integration
$1  << s << a^2$ or equivalently $a^{-2} << \bar s << 1$. Thus
we need to study the behaviour of  \refb{e6p}, \refb{e5q} for
small $\bar s$. Since both $K^s_{S^2}(0;s)$ and $K^s_{AdS_2}(0;s)$
diverge at $\bar 
s=0$, we cannot simply expand the summand /integrand
in \refb{e6p}, \refb{e5q} in a power series expansion in $\bar s$, 
-- we must first
isolate the divergent part and evaluate it exactly.
Let us begin with the expression for $K^s_{AdS_2}(0;s)$
given in \refb{e5q}. We first express $\tanh(\pi\lambda)$
as $1 -2\,  e^{-2\pi\lambda} / (1 + e^{-2\pi\lambda})
= 1 + 2\sum_{k=1}^\infty (-1)^k e^{-2k\pi\lambda}$, and divide the
integral into two parts: the first part containing the 1 term from
the expansion of $\tanh(\pi\lambda)$ and the second part
containing the rest of the terms. The first integral can be 
evaluated in closed form. In the second integral we
expand
$e^{-\bar s\lambda^2}$ in a power series expansion in $\bar s$
and perform the integral over $\lambda$. This leads to the
following expression for $K^s_{AdS_2}$:
\ben \label{e8}
K^s_{AdS_2}(0;s) &=& {1\over 4\pi a^2\, \bar s}\,
e^{-\bar s / 4} \, \left[ 1 + 
\sum_{n=0}^\infty {(-1)^n\over n!} (2n+1)! {\bar s^{n+1}
\over \pi^{2n+2}} {1\over 2^{2n}} \left(2^{-2n-1}-1\right)
\zeta(2n+2)\right]\nn
&=& {1\over 4\pi a^2\, \bar s}\,
e^{-\bar s / 4} \, \left(1 -{1\over 12} \bar s+ 
{7\over 480} \bar s^2 + \OO(\bar s^3) \right)\, .
\een
In order to find the small $s$ expansion of $K^s_{S^2}$,
we  first express \refb{e6p} as
\be \label{ess1}
{1\over 4\pi i\, a^2} \, e^{\bar s/4}\, 
\ointop \, d\lambda \, \lambda \, 
\tan(\pi\lambda)\, e^{-\bar s\lambda^2}\, ,
\ee
where $\ointop$ denotes integration along a contour that travels
from from $\infty$ to 0 staying below the real axis and returns to
$\infty$ staying above the real axis. By deforming the integration
contour to a pair of straight lines through the origin --
one at an angle $\kappa$ below the positive real
axis and the other at an angle $\kappa$ above the positive
real axis -- we can express this as
\be \label{ess2}
{1\over 2\pi a^2} e^{\bar s/4}
\, Im \, \int_0^{e^{i\kappa}\times \infty}
\, \lambda \, d\lambda\, \tan(\pi\lambda) \, e^{-\bar s\lambda^2}
\, , \qquad 0<\kappa<< 1\, .
\ee
This integral can now be expressed in the same way as in the case
of $K^s_{AdS_2}$, and we get
\ben \label{e9}
K^s_{S^2}(0;s) &=& {1\over 4\pi a^2\, \bar s}\,
e^{\bar s / 4} \, \left[ 1 -
\sum_{n=0}^\infty {1\over n!} (2n+1)! {\bar s^{n+1}
\over \pi^{2n+2}} {1\over 2^{2n}} \left(2^{-2n-1}-1\right)
\zeta(2n+2)\right]\nn
&=& {1\over 4\pi a^2\, \bar s}\,
e^{\bar s / 4} \, \left(1 +{1\over 12} \bar s+ 
{7\over 480} \bar s^2 + \OO(\bar s^3) \right)\, .
\een

Substituting \refb{e8} and \refb{e9} into \refb{e4a} we get
\be \label{e10}
K^s(0;s) =  
{1\over 16\pi^2 a^4\, \bar s^2} 
\left( 1 +{1\over 45} \bar s^2 +\OO(s^4)\right)\, .
\ee
Eq.\refb{e3ab} now gives
\be \label{e12}
\Delta\LL_{eff}= {1\over 32\pi^2 a^4} 
\, \int_{\eps/a^2}^\infty \, {d\bar s\over \bar s^3} \, 
\left( 1 +{1\over 45} \bar s^2 +\OO(\bar s^4)\right)\, .
\ee
This integral has a quadratically divergent piece proportional to
$1/\eps^2$. This can be thought of as a renormalization of the
cosmological constant and will cancel against 
contribution from other fields in a supersymmetric theory in which
the cosmological constant is not renormalized. 
Even otherwise in string theory there is a physical cut-off set
by the string scale.\footnote{Typically in a string theory there are
multiple scales {\it e.g.} string scale, Planck scale, scale set by
the mass of the D-branes etc. We shall consider near horizon
background where the string coupling constant as well as all the
other parameters describing the shape, size and the various background
fields along the six compact directions are of order unity. In this
case all these length scales will be of the same order. \label{f1}}
Our main interest is in
the logarithmically divergent piece which comes from the 
order $\bar s^2$ term inside the parentheses. This is given by
\be \label{e13}
{1\over 1440\pi^2 a^4} \, \ln (a^2/\eps)\, ,
\ee
and, according to \refb{e3} gives a contribution to the entropy
\be \label{e14}
\Delta S_{BH} = -{1\over 180} \ln (a^2 /\eps)\, .
\ee

This agrees with the earlier result of 
\cite{9408068,9709064}. In this
earlier approach one computed the black hole entropy by relating it
to the partition function of the theory in an eucldean space-time
with a conical defect\cite{9407001}.  
This required computing the scalar
heat kernel on a space-time with conical defect. Besides being
computationally more difficult, this method also suffered from
the intrinsic problem that string theory on a space-time with
conical defect may not be well defined. In contrast the
quantum entropy function approach only requires us to compute the
partition function of string theory on the near horizon $AdS_2\times
S^2$ geometry. Since this is a smooth geometry, and solves the
classical equations of motion of string theory, the partition function
of the theory in this space should be well defined, leading to an
unambiguous prescription for computing the black hole entropy.

Note that the term in $\LL_{eff}$ proportional to
$\ln a^2$ comes from the $s$ independent part of $K^s(0;s)$
in an expansion in $s$. This can also be calculated using
the general formula derived in \cite{dewitt, mckean,
christ-duff1,christ-duff2,gilkey,0306138} which relates the coefficients
appearing in the small $s$ expansion of $K(0;s)$ to local
quantities computed in the background geometry. 
In $AdS_2
\times S^2$, where the Weyl tensor as well as the
curvature scalar vanishes, the formula for the constant part
of $K^s(0;s)$, denoted as $a^s_4(0;s)$, takes the form
\be \label{elocal1}
a^s_4(0;s) = {1\over 180 (4\pi)^2} \, R_{\mu\nu} \,
R^{\mu\nu}
\, ,
\ee
where $R_{\mu\nu}$ is the Ricci scalar. Evaluating it on the background
\refb{e1} we find that $R_{\mu\nu}R^{\mu\nu}=4a^{-4}$, and
hence
\be \label{elocal2}
a^s_4(0;s) = {1\over 16\pi^2 a^4} \, {1\over 45}\, .
\ee
This is in precise agreement with the coefficient of the $s$ independent
term in \refb{e10}. We shall see however that evaluating the heat
kernel explicitly by summing over eigenfunctions gives us valuable
insight that will be useful for our analysis when we try to extend the
results to include the contribution from higher spin fields,
and the effect of background electric and magnetic fluxes
on $AdS_2\times S^2$. In particular in order that the integrals
of the form appearing in \refb{e3ab} are well defined, we need to
subtract from the integrand $K(0,s)$ its value at $s\to\infty$,
-- this corresponds to removing the zero eigenvalues of the
kinetic operator from the definition of the determinant. Thus we
need to know the $s\to\infty$ limit of the heat kernel besides
its small $s$ expansion.

Let us now discuss the effect of switching on a mass term for the
scalar. The effect of this is to insert a factor of $e^{-m^2 s}
= e^{-m^2 a^2 \bar s}$ into the integral in \refb{e12}.
This gives
\be \label{e15}
\Delta\LL_{eff}= {1\over 32\pi^2 a^4} 
\, \int_{\eps/a^2}^\infty \, {d\bar s\over \bar s^3} \, 
\left( 1 +{1\over 45} \bar s^2 +\OO(\bar s^4)\right)\, 
e^{-m^2 a^2 \bar s}\, .
\ee
We shall now consider two different situations. First suppose
$m^2$ is of order unity, \i.e.\ of the order of the string scale. In that
case the exponential factor in \refb{e15} effectively
restricts the integration over
$\bar s$ in the region $\bar s{<\atop \sim} 
1/a^2$. As a result $\Delta\LL_{eff}$ will
not contain any piece proportional to $\ln a^2$. On the other hand if
$m^2 = c / a^2$ where $c$ is a constant of order unity, then
in the region $\bar s << 1$ the term in the exponent is small and we can
expand the exponential in a Taylor series expansion. This gives
\be \label{e16}
\Delta\LL_{eff} =  {1\over 32\pi^2 a^4} 
\, \int_{\eps/a^2}^\infty \, {d\bar s\over \bar s^3} \, 
\left( 1 -c\bar s + {1\over 2}\,
c^2 \bar s^2+{1\over 45} \bar s^2 +\OO(\bar s^4)\right)\, .
\ee
This has a term proportional to $\ln a^2$ of the form
\be \label{e17}
{1\over 8\pi^2 a^4}\,
\left( {1\over 180} + {c^2\over 8}\right)
\, \ln (a^2)\, ,
\ee
and, according to \refb{e3} gives a contribution to the entropy
\be \label{e18}
\Delta S_{BH} = -\left({1\over 180} + {c^2\over 8}\right)
\ln (a^2)\, .
\ee
This shows that  a massive scalar whose mass is of the order of
string scale does not give any contribution to the entropy proportional
to $\ln a^2$. On the other hand a massive scalar whose mass is
inversely proportional to $a$ will contribute terms proportional to
$\ln a$ to the entropy, and furthermore the actual contribution will
depend on the mass of the scalar. 

In the examples we shall
analyze, the presence of the background flux generates
an effective potential of order $a^{-2}$ for background scalars 
via the coupling between scalars and vector fields in the
supergravity action. Thus we must take into account such corrections
in our analysis. In contrast the massive string modes have mass of the
order of the string scale and  we can ignore their
contribution while computing logarithmic correction to the black
hole entropy. Finally one might also worry about the effect of higher
derivative corrections to the effective action on the potential
for the scalars. Such corrections are of order $a^{-4}$ or higher
powers of $a^{-1}$, and  do not affect the logarithmic correction to the
entropy since correction to the exponent $m^2 a^2 \bar s$ 
remains small throughout the relevant region of integration.

So far we have discussed the effect of one loop corrections.
What about higher loop contributions? As discussed in footnote
\ref{f1}, we have assumed that all the moduli
parameters including the string coupling constant are of order
unity at the horizon; hence the higher loop contributions could
be of the same order as the one loop contribution. While we
cannot make any definite prediction about these higher loop
corrections in general, in the special case of supersymmetric
black holes we shall be considering, we can argue as
follows that the
higher loop contributions can be ignored. For definiteness we
shall consider a situation where all the charges carried by the
black hole are Ramond-Ramond (RR) charges. In this case
the scaling argument of \cite{0908.3402} tells us that as we scale
all the charges by some common scale $\Lambda$, the
dilaton $\Phi$ at the horizon scales as $e^{-2\Phi}\sim \Lambda^2$,
and all the other NSNS background fields, including the string metric,
remains fixed. 
As a result for large $\Lambda$ the four dimensional
canonical metric $g_{\mu\nu}$, related to the string metrc 
$G_{\mu\nu}$ via $g_{\mu\nu} = e^{-2\Phi}G_{\mu\nu}$, scales as
$\Lambda^2$ and $e^{-2\Phi}$ scales as $\Lambda^2$.
This would seem to contradict our
assumption of footnote \ref{f1} that all the moduli are of
order one at the horizon. This is resolved by
noting that at least in the classical supergravity approximation,
the value of the dilaton at the horizon can be changed keeping
the four dimensional
canonical metric fixed. In the language of $\NN=2$ supersymmetric
theories this is a consequence of the fact that the four dimensional
dilaton belongs to the hypermultiplet and hence the vector multiplet
fields do not generate any potential for the dilaton. We shall assume
that this flat direction, labelled by the value of the dilaton, 
is not lifted even in the full quantum theory.  Thus
we can evaluate the entropy at any value of the dilaton, in particular
either for $e^{-2\Phi}\sim \Lambda^2$ as given by the scaling
argument or for $e^{-2\Phi}\sim 1$. In the first case the
$k$ loop contribution to the entropy will go as $e^{2\Phi(k-1)}
\sim \Lambda^{2-2k}$.
In the second case we have $e^{2\Phi}\sim 1$
and $G_{\mu\nu}=e^{2\Phi} g_{\mu\nu}\sim \Lambda^2$.
Thus any term that has $2k+2$ derivatives
will give a contribution of order $\Lambda^{2-2k}$ irrespective of
the order of the perturbation theory 
in which it is generated. Requiring that
both be correct leads to the conclusion that at $k$ loop order only the
terms with $2k+2$ derivatives will contribute to the entropy,
giving a contribution of order $\Lambda^{2-2k}$.
Thus the logarithmic corrections can arise only from one loop
terms in the effective action, the higher loop corrections being
suppressed by inverse powers of $\Lambda$, \i.e.\ inverse
powers of the charges.

\renewcommand{\theequation}{\thesubsection.\arabic{equation}}

\sectiono{Heat kernels of  vector, $p$-form and fermion
fields} \label{sother}

The matter multiplet of an $\NN=4$ supersymmetric theory in
(3+1) dimensions contains a vector, four Majorana fermions and
six scalars. Thus in order to compute the logarithmic
correction to the entropy due to a matter multiplet we need to
extend the results of the previous section to include  the 
heat kernels of vector and fermion fields.
In this section we shall compute the heat kernels of these
fields by regarding them as free fields in $AdS_2\times S^2$
background. 
Although the analysis is straightforward using the results of
\cite{campo,camhig1,campo2,camhig2,9505009}, 
we shall go through it carefully, since, 
as we shall see in the next two sections, these
results need to be further corrected due to 
mixing between scalar, vector and tensor
fields in the black hole near horizon
geometry.

\subsectiono{Vector fields}

In general the contribution from a field of given spin
requires evaluation of the functional integral after suitable gauge
fixing. We shall use a Feynman type gauge and compute the 
net contribution from a given field 
as the
sum of the contribution from the original field as well as the various
ghosts which appear during gauge fixing. 
Let us first consider the case of a $U(1)$ gauge field with
euclidean action
\be \label{ea1}
\SSS_A = -{1\over 4} \int d^4 x \sqrt{\det g}
\, F_{\mu\nu} F^{\mu\nu}\, ,
\ee
where $F_{\mu\nu} \equiv \p_\mu A_\nu - \p_\nu A_\mu$ is the
gauge field strength. Adding a gauge fixing term
\be \label{ea2}
S_{gf} = -{1\over 2} \int d^4 x \sqrt{\det g} \, (D_\mu A^\mu)^2\, ,
\ee
we can express the action as
\be \label{ea3}
\SSS_A +\SSS_{gf}
= -{1\over 2} \int d^4 x \sqrt{\det g} A_\mu
(\Delta A)^\mu\, ,
\ee
where
\be \label{eddelta}
(\Delta \, A)_\mu \equiv -\square \, A_\mu + R_{\mu\nu}
A^\nu\, ,\qquad \square A_\mu \equiv g^{\rho\sigma} D_\rho D_\sigma
A_\mu\, .
\ee
We shall denote by $d$ the exterior derivative operator and by
$\delta$ the operator $-* d*$ where $*$ denotes Hodge dual operation.
Then $\Delta$ may be expressed as
\be \label{edefdelta}
\Delta\equiv
(d\delta + \delta d)\, .
\ee
We shall use \refb{edefdelta} as the definition of $\Delta$
acting on any $p$-form field.

Since the eigenfunctions of $\Delta$ are four component vectors,
the vector heat kernel is a $4\times 4$ matrix. We shall denote by
$K^v(x,x';s)$ the trace of this matrix. 
Quantization of gauge fields also requires us to introduce
two anticommuting scalar ghosts whose kinetic operator is given by
the standard laplacian $-\square=\delta d$ in the harmonic gauge.
Thus the net one loop contribution of the vector
field to $\LL_{eff}$ will be given by
\be \label{e19}
{1\over 2}\, \int_{\eps}^\infty\, {ds\over s} \, 
 \sqrt{\det g}\, \lim_{x'\to x} \, \left[
K^{v}(x,x';s) - 2\, K^s(x,x'; s) \right]\, ,
\ee
where the $-2\, K^s$ term reflects the contribution due to the
ghosts.

A vector in
$AdS_2\times S^2$ decomposes into a (vector, scalar) plus
a (scalar, vector), with the first and the second factors representing
tensorial properties in $AdS_2$ and $S^2$ respectively.
Furthermore, on any of these
components the action of the kinetic operator
can be expressed as $\Delta_{AdS_2}+\Delta_{S^2}$, with $\Delta$
as defined in \refb{edefdelta}. Thus
we can construct the eigenfunctions of $\Delta$
by taking the product of appropriate eigenfunctions of
$\Delta_{AdS_2}$ and $\Delta_{S^2}$, and the corresponding
eigenvalue of $\Delta$ on $AdS_2\times S^2$ will be given by the
sum of the eigenvalues of $\Delta_{AdS_2}$ and $\Delta_{S^2}$.
This gives\footnote{The main ingradient that
allows us to express the heat kernel on $AdS_2\times S^2$ in
terms of heat kernels on $AdS_2$ and $S^2$ is that the
kinetic operator on $AdS_2\times S^2$ can be expressed as
a sum of the kinetic operators in $S^2$ and $AdS_2$. 
This will continue to hold for the other fields as well, but
the choice
of harmonic gauge is essential for this.}
\be \label{e20}
K^{v}(0;s)=K^{v}_{AdS_2}(0,s) K^s_{S^2}(0;s)
+ K^s_{AdS_2}(0,s) K^{v}_{S^2}(0;s)\, .
\ee
Thus we need to compute $K^{v}_{AdS_2}(0,s)$ and
$K^{v}_{S^2}(0;s)$. 

 Now suppose that we have a scalar field $\Phi$ 
on $AdS_2$ or $S^2$
satisfying 
\be \label{e22}
 \Delta\Phi\equiv \delta d\, \Phi 
\equiv - \square \Phi   = \kappa\Phi\, .
\ee
Then we can construct two 
configurations for the gauge field $A$
with the same eigenvalue $\kappa$ of $\Delta$ and the same
normalization as $\Phi$ as follows:
\be \label{e23}
A^{(1)} = \kappa^{-1/2}\, d\Phi, \qquad  A^{(2)} = 
\kappa^{-1/2}\, * d\Phi\, .
\ee
Furthermore locally every vector field in
two dimensions can be decomposed as $d\Phi_1 + *d\Phi_2$.
Thus for every scalar eigenfunction $\Phi$
of the operator $\delta d$
we have a pair of vector eigenfunctions of $(d\delta + \delta d)$ with
the same eigenvalue. The  contribution from any of these two
eigenfunctions to the
vector heat kernel $K^v(x,x;s)$ is given by $\kappa^{-1}\,
e^{-\kappa s}\, 
g^{\mu\nu}\p_\mu\Phi(x)
\p_\nu\Phi(x)$. Now since $K^v(x,x;s)$ is independent
of $x$ after summing over the contribution from all the states, we could
compute it by taking the volume average of each term.
Taking a volume average 
allows us to integrate by parts and gives
the same result as the volume average of
$\kappa^{-1} \,
e^{-\kappa s}\, \Phi(x)\delta d \Phi(x)
= e^{-\kappa s}\, \Phi(x)^2$. 
This is the same as the contribution
from $\Phi(x)$ to the scalar heat kernel.\footnote{This can also be
verified using the explicit  form of the scalar eigenmodes given in
\S\ref{s2} and noting that the non-vanishing contribution now comes from
eigenmodes with $Y_{l,\pm 1}$ on $S^2$ and $f_{\lambda,\pm1}$ on
$AdS_2$.}
Thus we conclude that 
leaving aside global issues,
the heat kernel for a vector field on
$AdS_2$ or $S^2$ should be given by twice that of the scalar.

There are however some corrections to this both on $S^2$ and
$AdS_2$ due to global issues. 
On $S^2$, the constant mode of
the scalar is an eigenfunction of $\square_{S^2}$ 
with eigenvalue 0.
However these modes do not generate any non-trivial gauge field
configuration via \refb{e23}. Hence their contribution to 
$K^s_{S^2}$ should be removed while computing
$K^v_{S^2}$. Since the zero mode gives a
contribution of $1/(4\pi a^2)$ to $K^s_{S^2}(0;s)$, this gives
\be \label{e21a}
K^{v}_{S^2}(0,s) = 2\, K^s_{S^2}(0,s) 
- {1\over 2\pi a^2}\, .
\ee
On the other hand on $AdS_2$ the constant
mode of the scalar is not normalizable, and hence $K^s_{AdS_2}$
does not include any contribution from the constant mode. Thus
we do not need to  make any subtraction from $K^s_{AdS_2}$
in computing $K^v_{AdS_2}$. However it turns out that in this
case there is a set of square integrable
eigenvectors of $\Delta$ with zero
eigenvalue, given by\cite{camhig1}:\footnote{Since $d\Phi$ is
(anti-)self-dual in $AdS_2$, we do not get independent eigenfunctions
from $*d\Phi$.}
\be \label{e24}
A = d\Phi, \qquad \Phi = {1\over \sqrt{2\pi |\ell|}}\,
\left[ {\sinh\eta \over 1+\cosh\eta}\right]^{|\ell|} e^{i\ell\theta},
\quad \ell = \pm 1, \pm 2, \pm 3, \cdots\, .
\ee
These are not included in \refb{e23} since the $\Phi$ given in
\refb{e24} is not normalizable. These give additional contribution
to $K^v_{AdS_2}(0;s)$. In fact since for $|\ell|>1$ the gauge
field vanishes at $\eta=0$, only the $\ell=\pm 1$ terms contribute
to $K^v_{AdS_2}(0;s)$. 
This gives
\be \label{e21b}
K^{v}_{AdS_2}(0,s) = 2\, K^s_{AdS_2}(0,s) 
+ {1\over 2\pi a^2}\,  .
\ee

We now proceed to compute the contribution to 
the vector heat kernel 
using these results. Using  \refb{e20}, \refb{e21a}, \refb{e21b} 
and then \refb{e8}, \refb{e9}
we get
\ben \label{e25}
K^v(0;s) &=& 4 K^s_{AdS_2}(0;s) K^s_{S^2}(0;s) +{1\over
2 \pi a^2}
\left(K^s_{S^2}(0;s) - K^s_{AdS_2}(0;s)
\right)  \nn
&=& 
{1\over 4\pi^2 a^4 \bar s^2} \left(1 +{16\over 45}
\bar s^2 +\OO(\bar s^4)\right)\, .
\een
This is again consistent with the results of \cite{christ-duff1}
which gives the coefficient of the $s$ independent part of
$K^v$ to be $64 (180)^{-1} (4\pi)^{-2} 
R_{\mu\nu} R^{\mu \nu}$. Taking into account the 
contribution due to the ghosts via
\refb{e19} we can now compute the total contribution to the
effective action from the vector field.

There is however an additional subtlety we must take care of.
The contribution to the vector heat kernel given in
\refb{e25} includes contribution from the zero modes obtained
by taking the product of \refb{e24} and the $l=0$ mode,
\i.e.\ the constant mode of the
scalar on $S^2$. The
integration over the zero modes of any
field requires special treatment since these integrals are not Gaussian.
Thus in evaluating the determinant of the kinetic operator for
computing the one loop contribution to the effective action we must
remove the contribution due to the zero modes\cite{9505186}.
This will require replacing $K^v(0;s)$ by
\be \label{e20a}
\wh K^{v}(0;s)=K^{v}_{AdS_2}(0,s) K^s_{S^2}(0;s)
+ K^s_{AdS_2}(0,s) K^{v}_{S^2}(0;s) - {1\over 8\pi^2 a^4}\, .
\ee
More generally, removal of the zero modes from any heat
kernel will require subtracting from  $K(0;s)$ its 
value as $s\to\infty$:
\be \label{edefkp}
\wh K(0;s) \equiv K(0;s) - \lim_{t\to\infty} K(0,t)\, .
\ee
We shall take this as the definition of the proper
heat kernel that should be used in computing logrithmic correction to
the entropy.
This subtraction is in fact necessary to ensure that the integration
over $s$ does not diverge at 
infinity.\footnote{In any case a constant term in
$K(0;s)$ will not produce a factor of $\ln a^2$, -- these arise
from terms which remain constant in the range 
$1<<s<< a^2$ and fall off for $s>>a^2$.}
However instead of removing the zero mode 
contribution from
the heat kernel of every field we shall find it more convenient to
remove the contribution at the end from the trace
of the total heat kernel
of all the fields. For this reason we shall continue to use the
result \refb{e25} for the vector heat kernel.

Even though in evaluating the determinant of the kinetic operator we
need to remove the contribution due to the zero modes, eventually we
must carry out the integration over the zero modes of physical
fields.
We shall describe the analysis of the zero mode integrals in
appendix \ref{sa} and show that the net effect of these integrals --
and an additional contribution that will be described in the same 
appendix --
cancel, leaving us with the prescription of working with the
regularized heat kernel described in \refb{edefkp}.

\subsectiono{$p$-form fields} 

A matter multiplet in $\NN=4$ supergravity theory contains 
 six scalar fields. However often
in string theory, some of the scalars appear in their dual form as
2-form fields. This happens for example if we consider type IIA
string theory on $K3\times T^2$, -- we get 2-form fields from taking the
components of the RR 3-form field with one leg on $T^2$ and also
from the NSNS sector 2-form fields. All of these need to be dualized
to scalars and they then
form parts of the matter multiplets. However from
the viewpoint of type IIA string theory we should really carry out the
path integral by regarding them as 2-form fields. Similarly in the
same theory the RR
3-form field with all its legs along the four dimensional
Minkowski space must also be regarded as an integration variable
in the path integral even though in four dimensions it does not have
any physical degree of freedom. Thus in order that our results do not
depend on which description of the theory we use, we must ensure that the
2-form field and the scalar gives the same contribution to the
one loop determinant and that the 3-form field does not contribute
to the one loop determinant. We shall now try to verify this
explicitly. This analysis is important in view of the results of
\cite{duffnieu} that the dual descriptions do not always lead to the same
result for the trace of the stress tensor, which in turn can be
related to the $s$ independent term in the expansion of
$K(0;s)$.

First we consider the 2-form field $B_{\mu\nu}$ with gauge
invariant action
\be \label{e2f1}
S_B = -{1\over 12} \, \int d^4 x\, \sqrt{\det g} \, H_{\mu\nu\rho}
H^{\mu\nu\rho}, \qquad H_{\mu\nu\rho} =\p_\mu B_{\nu\rho}
+\p_\nu B_{\rho\mu} + \p_\rho B_{\mu\nu}\, .
\ee
Adding a harmonic gauge fixing term
\be \label{e2f2}
S_{gf} = -{1\over 2} \int d^4 x\, \sqrt{\det g}\, g^{\mu\nu} \,
D^\rho\, B_{\rho\mu} \, D^\sigma \, B_{\sigma\nu}\, ,
\ee
we get  a simple form of the total action
\be \label{e2f3}
S_B + S_{gf} = -{1\over 2}\, \int d^4 x\,
\sqrt{\det g} \, B_{\mu\nu} 
(\Delta B)^{\mu\nu}\, , \quad \Delta B
\equiv (d\delta + \delta d)B\, .
\ee
On $AdS_2$ and
$S^2$, there is a one to one correspondence between the normalizable
modes of the scalar and the normalizable modes of $B_{\mu\nu}$ via
Hodge duality $B_{\mu\nu}=\Phi\, \vareps_{\mu\nu}$  where
$\vareps$ is the solume form.  
As a result the heat kernels for $B_{\mu\nu}$ and scalars
are identical in these two spaces:
\be \label{e29}
K^b_{S^2}(0;s) = K^s_{S^2}(0;s), \qquad
K^b_{AdS_2}(0;s) = K^s_{AdS_2}(0;s)\, .
\ee
Since the 2-form field on $AdS_2\times S^2$ can be decomposed as
(vector, vector), (2-form,  scalar) and (scalar, 2-form), and 
furthermore on any of these components
the action of the kinetic operator is given by $\Delta_{AdS_2}
+\Delta_{S^2}$, we can express the 
trace of the heat kernel of the
2-form field on $AdS_2\times S^2$ as
\ben \label{e30}
K^b(0;s) &=& K^v_{AdS_2}(0;s) K^v_{S^2}(0;s) +
2\, K^s_{AdS_2}(0;s) \, K^s_{S^2}(0;s) \nn
&=& 6\, K^s_{AdS_2}(0;s) K^s_{S^2}(0;s) 
+ {1\over \pi a^2}
\left(K^s_{S^2}(0;s) - K^s_{AdS_2}(0;s)\right) - {1\over 4\pi^2 a^4}
\nn
&=& {1\over 16\pi^2 a^4} \left( {6\over \bar s^2}-{6\over 5}
+\OO(\bar s^2)\right)
\, ,
\een
where we have used \refb{e29} in the first step,  \refb{e21a},
\refb{e21b} in
the second step and \refb{e8}, \refb{e9} in the last step.
This is consistent with the results of \cite{christ-duff1}
which gives the coefficient of the $s$ independent part of
$K^b$ to be $-54 (180)^{-1} (4\pi)^{-2} 
R_{\mu\nu} R^{\mu \nu}$.

Quantization of the 2-form field produces two anti-commuting
vector ghosts and three commuting scalar 
ghosts\cite{siegel,ThierryMieg,sezgin}. 
Thus the net
contribution to the one loop effective action is given by
\ben \label{e31}
\Delta \LL_{eff} &=& {1\over 2} \, \int_{\eps}^\infty
{d s\over  s} \left[K^b(0;s) - 2\, K^v(0;s)
+ 3 K^s(0;s)\right] \nn
&=& {1\over 2} \, \int_{\eps}^\infty
{d s\over  s} \left[ K^s(0;s) - {1\over 4\pi^2 a^4}\right]
\, .
\een
Comparing \refb{e3ab} and \refb{e31} we see that the contribution
to the one loop effective action due to a scalar field differs from that
of a 2-form field. 
This is a bit surprising since in four dimensions the
scalar and 2-form fields are supposed to be equivalent.
As already noted in
\cite{duffnieu,0806.3505}, this difference can be attributed to the
contribution due to the zero modes, -- we shall now verify this
explicitly. Indeed the zero mode contribution to the heat kernel
can be identified as the term obtained by taking the $s\to \infty$
limit of the heat kernel. Thus on the right hand side of
\refb{e31} this is given by the $- 1/ 4\pi^2 a^4$ term in
the square bracket. 
As discussed before, in calculating the one loop determinant
we must explicitly remove the contribution due to the
zero modes. In this case 
we shall no longer have the 
$- {1/4\pi^2 a^4}$ term inside the integrand in \refb{e31} and the
result for the effective action computed using the 2-form field would
agree
with that computed using the scalar.

We can
carry out a similar analysis for a 3-form field in the harmonic gauge.
In this gauge the kinetic operator is again given by $\Delta 
=(d\delta + \delta d)$. 
Since the 3-form
field on $AdS_2\times S^2$ and a vector field can be related by
Hodge duality, the relevent part
of the heat kernel for the 3-form is given by
$K^v(0;s)$. 
On the other hand the quantization of the 3-form
requires 2 anti-commuting 2-form ghosts, 3 
commuting vector ghosts and 4 anti-commuting 
scalar ghosts\cite{siegel,ThierryMieg,sezgin}.
Thus the net contribution to $\Delta\LL_{eff}$ is
\ben \label{e34}
\Delta \LL_{eff} &=& {1\over 2} \, \int_{\eps}^\infty
{d s\over  s} \left[K^v(0;s) - 2\, K^b(0;s)
+ 3 K^v(0;s) - 4 K^s(0;s)\right] \nn
&=& 
{1\over 4\pi^2 a^4} \, \int_{\eps}^\infty
{d s\over  s}\, .
\een
This is contrary to our expectation that the contribution to the effective
action from a 3-form field should vanish since it is non-dynamical in
four dimensions. 
We now note that since the total heat kernel represented by
the term inside the square bracket is an $s$-independent constant,
removing the zero mode contribution amounts to subtracting this
constant.
This makes the net contribution vanish, in agreement with the
general expectation.

\subsectiono{Fermions}

Next we turn to the computation of the heat kernel of 
spinors\cite{9505009}.
Consider a Dirac spinor\footnote{Even if the spinors satisfy Majorana/Weyl
condition, we shall compute their heat kernel by first computing the result
for a Dirac spinor and then taking appropriate square roots.}
on $AdS_2\times S^2$. It decomposes
into a product of a Dirac spinor on $AdS_2$ and a Dirac spinor
on $S^2$. 
We use the following conventions for the vierbeins and the gamma
matrices
\be \label{evier1}
e^0= a\, \sinh\eta \, d\theta, \quad e^1 = a\, d\eta, \quad
e^2 = a\, \sin\psi\, d\phi, \quad e^3 = a\, d\psi\, ,
\ee
\be \label{egam1}
\Gamma^0 = -\sigma_3\otimes \tau_2, \quad \Gamma^1 = \sigma_3
\otimes \tau_1, \quad \Gamma^2 = -\sigma_2\otimes I_2, \quad 
\Gamma^3 = \sigma_1\otimes I_2\, ,
\ee
where $\sigma_i$ and $\tau_i$ are two dimensional Pauli matrices
acting on different spaces and $I_2$ is $2\times 2$ identity
matrix. In this convention 
the Dirac operator on
$AdS_2\times S^2$ can be written as
\be \label{ek1}
\not \hskip-4pt D_{AdS_2\times S^2}
= \not \hskip-4pt D_{S^2} + \sigma_3 \, 
\not \hskip-4pt D_{AdS_2}\, ,
\ee
where
\be \label{ed1}
\not \hskip -4pt D_{S^2} = a^{-1}\left[ -
\sigma^2\, {1\over \sin\psi} \p_\phi
+ \sigma^1 \, \p_\psi +{1\over 2}\, \sigma^1\, \cot\psi\right]\, ,
\ee
and
\be \label{ed1a}
\not \hskip -4pt D_{AdS_2} =
a^{-1}\left[ -\tau^2\, {1\over \sinh\eta} \p_\theta
+ \tau^1 \, \p_\eta +{1\over 2}\, \tau^1\, \coth\eta\right]\, .
\ee

First let us analyze the eigenstates of $\not\hskip -4pt D_{S^2}$.
They
are given by\cite{9505009}
\ben \label{ed2}
\chi_{l,m}^{\pm} &=& {1\over \sqrt{4\pi a^2}}\,
{\sqrt{(l-m)!(l+m+1)!}\over l!}\,
e^{ i\left(m+{1\over 2}\right)\phi} 
\pmatrix{ i\, \sin^{m+1}{\psi\over 2}\cos^m {\psi\over 2}
P^{\left(m+1, m\right)}_{l-m}(\cos\psi)
\cr
\pm \sin^{m}{\psi\over 2}\cos^{m+1} {\psi\over 2}
P^{\left(m, m+1\right)}_{l-m}(\cos\psi)
}, \nn
\eta_{l,m}^{\pm} &=& {1\over \sqrt{4\pi a^2}}\,
{\sqrt{(l-m)!(l+m+1)!}\over l!}\,
e^{ -i\left(m+{1\over 2}\right)\phi} 
\pmatrix{ \sin^{m}{\psi\over 2}\cos^{m+1} {\psi\over 2}
P^{\left(m, m+1\right)}_{l-m}(\cos\psi)
\cr
\pm i \, \sin^{m+1}{\psi\over 2}\cos^m {\psi\over 2}
P^{\left(m+1, m\right)}_{l-m}(\cos\psi)}, 
\nn
&& 
\qquad l,m\in \ZZZ, \quad l\ge 0, \quad 0\le m\le l\, ,
\een
satisfying
\be \label{ed3}
\not \hskip -4pt D_{S^2} \chi_{l,m}^\pm =\pm i\, a^{-1}\,
\left(l +1\right) 
\chi_{l,m}^\pm\, , \qquad
\not \hskip -4pt D_{S^2} \eta_{l,m}^\pm =\pm i\, a^{-1}\,
\left(l +1\right) 
\eta_{l,m}^\pm\, .
\ee
Here $P^{\alpha,\beta}_n(x)$ are the Jacobi Polynomials:
\be \label{ed4}
P_n^{(\alpha,\beta)}(x) = { (-1)^n\over 2^n \, n!} (1-x)^{-\alpha}
(1+x)^{-\beta} {d^n\over dx^n} \left[ (1-x)^{\alpha+n}
(1+x)^{\beta+n}\right]\, .
\ee

We shall denote by $K^f_{S^2}(x,x';s)$ the trace over the spinor
indices of the heat kernel of the Dirac fermion on $S^2$. The
precise normalization of $K^f_{S^2}$ is chosen as follows.
If $\not \hskip -4pt D_{S^2}$ has 
eigenfunction $f_n(x)$  with eigenvalue $i\lambda_n$,
then we define
\be \label{ed5}
K^f_{S^2}(x,x';s) =- \sum_n e^{-s\lambda_n^2} 
f_n^\dagger(x') f_n(x)\, .
\ee
The extra minus sign in the definition of $K^f_{S^2}$ has been
included to account for the fact that for fermionic path integral we get
a factor of the determinant instead of the inverse of the determinant.
Two additional normalization factors cancel;  the fact that 
$i\not \hskip -4pt D$ is the square root of $-\not \hskip -4pt D^2$ gives
a factor of 1/2, but since we are considering a Dirac fermion instead
of a Majorana fermion we get a factor of 2.
The result for \refb{ed5}
in the $x\to x'$ limit can be simplified by noting
that for $\psi=0$, $\chi^\pm_{l,m}$, $\eta^\pm_{l,m}$ vanishes
unless $m=0$, and
\be \label{ed5a}
\left(\chi^\pm_{l,0}\right)^\dagger \chi^\pm_{l,0}
= \left(\eta^\pm_{l,0}\right)^\dagger \eta^\pm_{l,0}
={1\over {4\pi a^2}}\, (l+1)\, .
\ee
Thus we get
\be \label{ed7}
K^f_{S^2}(0;s) = -{1\over 2\pi \, a^2} \sum_{l=0}^\infty
(2l+2)
\, e^{-s\left(l+1\right)^2/a^2}\, .
\ee

The eigenstates of $\not\hskip -4pt D_{AdS_2}$
are given by the analytic continuation of the
eigenfunctions given in \refb{ed2}\cite{9505009}, making the
replacement $\psi\to i\eta$, $l\to -i\lambda -1$,
$\phi\to\theta$,
\ben \label{ed2a}
\chi_{m}^{\pm}(\lambda) &=& {1\over \sqrt{4\pi a^2}}\,
\left|{\Gamma\left( {1} + m + i\lambda\right)
\over \Gamma(m+1) \Gamma\left({1\over 2}+i\lambda\right)}\right|\,
e^{ i\left(m+{1\over 2}\right)\theta}  \nn
&& \qquad 
\pmatrix{ i \, {\lambda\over m+1}\, 
\cosh^{m}{\eta\over 2}\sinh^{m+1} {\eta\over 2}
F\left(m+1+i\lambda, m+1-i\lambda; m+2;-\sinh^2{\eta\over 2}\right)
\cr
\pm \cosh^{m+1}{\eta\over 2}\sinh^m {\eta\over 2}
F\left(m+1+i\lambda, m+1-i\lambda; m+1;-\sinh^2{\eta\over 2}\right)}, \nn \cr \cr
\eta_{m}^{\pm}(\lambda) &=& {1\over \sqrt{4\pi a^2}}\,
\left|{\Gamma\left( {1} + m + i\lambda\right)
\over \Gamma(m+1) \Gamma\left({1\over 2}+i\lambda\right)}\right|\,
e^{ -i\left(m+{1\over 2}\right)\theta}\nn
&&  \qquad 
\pmatrix{ \cosh^{m+1}{\eta\over 2}\sinh^m {\eta\over 2}
F\left(m+1+i\lambda, m+1-i\lambda; m+1;-\sinh^2{\eta\over 2}\right)
\cr
\pm i \, {\lambda\over m+1}\, 
\cosh^{m}{\eta\over 2}\sinh^{m+1} {\eta\over 2}
F\left(m+1+i\lambda, m+1-i\lambda; m+2;-\sinh^2{\eta\over 2}\right)
}, 
\nn \cr && 
\qquad  m\in \ZZZ, \quad 0\le m<\infty, \quad 0<\lambda<\infty\, ,
\een
satisfying
\be \label{eadsev}
\not \hskip -4pt D_{AdS_2} \chi_{m}^{\pm}(\lambda)=\pm i\, a^{-1}\,
\lambda\,  \chi_{m}^{\pm}(\lambda)
\, , \qquad
\not \hskip -4pt D_{AdS_2} \eta_{m}^{\pm}(\lambda)
=\pm i\, a^{-1}\,
\lambda\,  \eta_{m}^{\pm}(\lambda)
\, .
\ee
This gives
\ben \label{kfads}
K^f_{AdS_2}(0;s)
&=& -
\int_0^\infty d\lambda e^{-s\lambda^2 / a^2} \nn
&&
\sum_{m=0}^\infty \left[
(\chi_{m}^{+}(\lambda))^\dagger \chi_{m}^{+}(\lambda)
+(\chi_{m}^{-}(\lambda))^\dagger \chi_{m}^{-}(\lambda)
+(\eta_{m}^{+}(\lambda))^\dagger \eta_{m}^{+}(\lambda)
+(\eta_{m}^{-}(\lambda))^\dagger \eta_{m}^{-}(\lambda)
\right]\nn
&=& -{1\over \pi a^2} \int_0^\infty d\lambda e^{-\bar s\lambda^2}
\, \lambda \, \coth(\pi\lambda)\, .
\een
In arriving at \refb{kfads} we have evaluated $\chi_m^\pm(\lambda)$,
$\eta_m^\pm(\lambda)$ at $\eta=0$ since the final result is
independent of the point in $AdS_2$ where we evaluate it.

The expansion of $K^f_{S^2}(s;0)$ and $K^f_{AdS_2}(s;0)$ 
for small $s$ can be found in the same way as for $K^s_{S^2}$ and
$K^s_{AdS_2}$. We get
\ben \label{esmallads}
K^f_{AdS_2}(0;s) 
&=& -{1\over 2\pi a^2\, \bar s}\,
\left[ 1 +
\sum_{n=0}^\infty {(-1)^n\over n!} (2n+1)! {\bar s^{n+1}
\over \pi^{2n+2}} {1\over 2^{2n}} 
\zeta(2n+2)\right]\nn
&=& -{1\over 2\pi a^2\, \bar s}\,
\left(1 +{1\over 6} \bar s-
{1\over 60} \bar s^2 + \OO(\bar s^3) \right)\, ,
\een
and
\ben \label{esmalls}
K^s_{S^2}(0;s) &=& -{1\over 2\pi a^2\, \bar s}\,
\left[ 1 -  \sum_{n=0}^\infty {1\over n!} (2n+1)! {\bar s^{n+1}
\over \pi^{2n+2}} {1\over 2^{2n}} 
\zeta(2n+2)\right]\nn
&=& -{1\over 2\pi a^2\, \bar s}\,
\left(1 -{1\over 6} \bar s - 
{1\over 60} \bar s^2 + \OO(\bar s^3) \right)\, .
\een

Now suppose that
$\psi_1$ denotes an eigenstate of $\not \hskip-4pt D_{S^2}$
with eigenvalue $i\tk_1$ and
$\psi_2$ denotes an eigenstate of $\not \hskip-4pt D_{AdS_2}$
with eigenvalue $i\tk_2$:
\be \label{ek2}
\not \hskip-4pt D_{S^2} \psi_1 = i\tk_1 \, \psi_1,
\qquad \not \hskip-4pt D_{AdS_2} \psi_2 = i\tk_2 \, \psi_2.
\ee
Since $\sigma_3$ anti-commutes with 
$\not \hskip-4pt D_{S^2}$ and commutes with 
$\not \hskip-4pt D_{AdS_2}$, we
have, using \refb{ek1},
\ben \label{ek3}
\not \hskip-4pt D_{AdS_2\times S^2}\, \psi_1\otimes \psi_2
&=&  i\tk_1 \psi_1\otimes \psi_2 +
i\tk_2 \sigma_3 \, \psi_1\otimes \psi_2 \, , \nn
\not \hskip-4pt D_{AdS_2\times S^2} \, \sigma_3\,
\psi_1\otimes \psi_2
&=& i\tk_2  \, \psi_1\otimes \psi_2 
 -i\tk_1 \sigma_3 \, \psi_1\otimes \psi_2
\, .\nn
\een
Diagonalizing the $2\times 2$ matrix we see that
$\not \hskip-4pt D_{AdS_2\times S^2}$ has eigenvalues
$\pm i\sqrt{\tk_1^2 + \tk_2^2}$. Thus the square
of the eigenvalue of $\not \hskip-4pt D_{AdS_2\times S^2}$ is
given by the sum of squares of the eigenvalues of
$\not \hskip-4pt D_{AdS_2}$ and $\not \hskip-4pt D_{S^2}$.
This in turn gives
\be \label{ek4}
K^f_{AdS_2\times S^2} = -K^f_{AdS_2} \, K^f_{S^2}\, ,
\ee
where the minus sign again accounts for the fact that the fermionic integration
produces a factor of the determinant instead of the inverse of the
determinant.
Using \refb{esmallads} and \refb{esmalls} we get
\be \label{ek5}
K^f_{AdS_2\times S^2} = -{1\over 4\pi^2 a^4\, \bar s^2} 
\left( 1 -{11\over 180} \, \bar s^2 + O(\bar s^3) \right)\, .
\ee
The $s$ independent term
in this expression 
is in agreement with the results of \cite{christ-duff1}.

\sectiono{Effect of graviphoton background in $\NN=4$
supersymmetric string theory} \label{sflux}

Quarter BPS black holes in $\NN=4$ supersymmetric string theories,
obtained by compactifying heterotic string theory on $T^6$ or 
equivalently type II string theory on $K3\times T^2$, have near
horizon $AdS_2\times S^2$ geometry. The background is also
accompanied by flux of electromagnetic fields along $AdS_2$ and
$S^2$. The presence of this flux modifies the kinetic terms of various
fields around this background, and hence also the associated
heat kernels. In this section we shall compute the modification of
the kinetic term
of various fields due to these fluxes.

\subsectiono{Four dimensional $\NN=4$ supergravity from ten
dimensional $\NN=1$ supergravity}

We shall begin by reviewing the dimensional reduction of the
ten dimensional supergravity action on $T^6$ leading to the four
dimensional $\NN=4$ supergravity action.\footnote{We could
have directly began with the $\NN=4$ supergravity action in
four dimensions given in \refb{1.3b}. However for dealing with
the fermions we have found it more convenient to use the
ten dimensional description.}
The
action of $\NN=1$ supergravity in ten dimensions coupled to
16 Maxwell fields is given by\cite{GSW}:
\ben \label{new1}
&& {1\over (2\pi)^6 (\alpha')^{4}} \int d^{10}z \sqrt{ \det
G^{(10)}}\,
e^{-2\Phi^{(10)}}\bigg[R^{(10)} +{1\over 4}
G^{\ten MN}\p_M\Phi^\ten \p_N\Phi^\ten
\nonumber \\
&& - {1\over 12} H^{(10)}_{MNP}
H^{(10)MNP} - {1\over 4} F^{(10)I}_{MN} F^{(10)IMN}
\nn
&& -\bigg\{
{1\over 2} 
\bar \psi^{\ten}_M \Gamma^{MNP} D_N \psi^{\ten}_P
+{1\over 2} \bar \Lambda \Gamma^M D_M \Lambda 
+{1\over 2} \bar \Sigma^I \Gamma^M D_M \Sigma^I 
\nn
&& -{1\over 48}
\bar \Sigma^I \Gamma^{MNP} \Sigma^I H^\ten_{MNP} 
+ {1\over 2\sqrt 2} \bar\Sigma^I \Gamma^M \Gamma^{NP}
\left(\psi^{\ten}_M + {1\over 6\sqrt 2} \Gamma_M\Lambda
\right) F^{\ten I}_{NP}
\nn && 
-{1\over 48} \Big( \bar \psi^{\ten}_M \Gamma^{MNPQR}
\psi^{\ten}_R + 6\bar\psi^{\ten N} \Gamma^P \psi^{\ten Q} 
- \sqrt 2 \bar\psi^{\ten}_M
\Gamma^{NPQ} \Gamma^M \Lambda \Big)
H^\ten_{NPQ} \Big\} \bigg]\nn
&& +\cdots\, .
\een
Here $G^{(10)}_{MN}$, $B^{(10)}_{MN}$, $A^{(10)I}_M$, and
$\Phi^{(10)}$ are ten dimensional metric, anti-symmetric tensor
field, U(1) gauge fields and the scalar dilaton field
respectively ($0\le M, N \le 9$, $1\le I\le 16$), 
$\psi^{\ten}_M$ denotes a left-handed Majorana-Weyl 
gravitino field, $\Lambda$ is a right-handed Majorana-Weyl
spinor field and
$\Sigma^I$ are left-handed Majorana-Weyl spinor 
fields in the gauge multiplet.
$\cdots$ denotes terms containing fermion bilinears multiplied
by derivatives of the dilaton or terms quartic in the fermions, and
\ben \label{new2}
F^{(10)I}_{MN} &=& \p_M A^{\ten I}_N - \p_N 
A^{\ten I}_M \nonumber\\
H^{(10)}_{MNP} &=& (\p_M B^\ten_{NP} -{1\over 2} A_M^{\ten I}
F^{\ten I}_{NP}) + \hbox{cyclic permutations in $M$, $N$, $P$},
\een
\ben \label{edefcov}
 D_M \psi^{(10)}_P &=& \p_M \psi^{(10)}_P - 
 \left\{{N\atop M~P}\right\} \psi^{(10)}_N
+{1\over 4} \, \omega^{AB}_M \Gamma^{AB} \psi^{(10)}_P\, , \nn
D_M \pmatrix{\Sigma^I\cr \Lambda} &=& \left( \p_M 
+{1\over 4} \, \omega^{AB}_M \Gamma^{AB}\right) 
\pmatrix{\Sigma^I\cr \Lambda}, \nn 
\omega^{AB}_M &=& -G^{(10)NP} e_N^{~B} \p_M e_P^{~A} +
e_N^{~A} e_P^{~B} G^{(10)PQ} \left\{{N\atop Q~M}\right\}, \nn
\left\{{M\atop N~P}\right\} 
&=& {1\over 2} G^{(10)MR} 
\left(\p_N G^{(10)}_{PR}
+\p_P G^{(10)}_{NR} - \p_R G^{(10)}_{NP}\right)\, ,
\een
the $e_M^{~A}$ being the vielbeins.
$\Gamma^A$'s are the 
$32\times 32$
$SO(10)$ gamma matrices,
$\Gamma^{AB}\equiv 
(\Gamma^A \Gamma^B
-\Gamma^B\Gamma^A)/2$, and
\be \label{emajo}
\bar \psi^{(10)}_M \equiv\psi^{(10)T}_M\, C, \quad 
\bar\Lambda \equiv
\Lambda^T\, C, \quad \bar \Sigma^I \equiv \Sigma^{IT}\, C\, ,
\ee
where $T$ denotes transpose
and $C$ is the $SO(10)$ charge conjugation matrix satisfying
\be \label{echco}
(C\Gamma^A)^T = C\Gamma^A\, .
\ee
We can use the vielbeins to convert
the tangent space indices to coordinate indices and vice versa.
We shall use the same symbol $\Gamma$ for labelling the
gamma matrices carrying coordinate indices. 
Our choice for the ten dimensional gamma matrices
and the charge conjugation matrix will be
as follows:
\ben \label{egamma}
&& \Gamma^0 = -\sigma_3\otimes \tau_2\otimes I_8, \quad 
\Gamma^1 = \sigma_3
\otimes \tau_1\otimes I_8, \quad \Gamma^2 = -
\sigma_2\otimes I_2\otimes I_8, \quad
\Gamma^3 = \sigma_1\otimes I_2\otimes I_8, \nn
&&
\Gamma^{p} = \sigma_3\otimes \tau_3\otimes
\wh \Gamma^{p}\, , \quad
C=\sigma_2\otimes \tau_1\otimes \wh C,
\quad 4\le p\le 9\, ,
\een
where $\wh\Gamma^{p}$ are $8\times 8$
$SO(6)$ gamma matrices and $\wh C$ is the
$SO(6)$ charge conjugation matrix satisfying
\be \label{esogamma}
\{\wh\Gamma^{p}, \wh\Gamma^{q}\} = 
2\delta_{{p}{q}}, \qquad (\wh C\wh\Gamma^{p})^T 
=-\wh C\wh\Gamma^{p}, \qquad \wh C^T =\wh C\, .
\ee
The spinors are
taken to be 32 component, and the Weyl condition is imposed by
setting to zero half of these components.
Finally note that in ten euclidean
dimensions we cannot impose Majorana-Weyl
condition and hence must formally
allow the
spinors to be complex. However we shall continue to use
\refb{emajo} as the definition of the barred fields. As a result
the action will not be real.

The supersymmetry transformation laws of various fields can be found
in the standard literature (see {\it e.g.} \cite{GSW}). We shall
only need to know the supersymmetry transformation laws of 
$\psi^{(10)}_M$ and $\Lambda$. 
In the background where all scalar fields are
constants this has the form:
\be \label{esusygrav}
\delta\psi^{(10)}_M = D_M \eta +{1\over 96} \left(\Gamma_M^{~NPQ} -
9 \delta_M^N \Gamma^{PQ}\right) \, H_{NPQ}\, \eta\, ,
\qquad \delta\Lambda = {1\over 24\sqrt 2}
\Gamma^{MNP}H_{MNP}\, \eta\, ,
\ee
where $\eta$ is the supersymmetry transformation parameter.

For dimensional reduction, it is convenient to introduce the
`four dimensional fields'  $\wh G_{\check a\check 
b}$, $\wh B_{\check a\check b}$, $\wh
A^I_{\check a}$, $\Phi$, $A_\mu^i$, 
$G_{\mu\nu}$ and $B_{\mu\nu}$ for 
$4\le \check a, \check b \le 9$, 
$0\le \mu, \nu \le 3$, $1\le I \le 16$ and $1\le i\le 28$
through the
relations\cite{9207016,9402002}
\ben\label{1.2}
&& \wh G_{\check a\check 
b}  = G^\ten_{\check a\check b }, \quad  \wh B_{\check a
\check b}  =
B^\ten_{\check a\check b}, \quad  \wh A^I_{\check a}
  = A^{\ten I}_{\check a},
\nonumber \\
&& A^{\check a-3}_\mu  = {1\over 2}\wh G^{\check a
\check b} G^\ten_{\check b \mu}, \quad
A^{I+12}_\mu = -({1\over 2} A^{\ten I}_\mu - \wh A^I_{\check b}
A^{\check b-3}_\mu), \nonumber \\
&&  A^{\check a+3}_\mu = {1\over 2}
B^\ten_{\check a \mu} - \wh B_{\check a\check b} A^{
\check b-3}_\mu 
+ {1\over 2}\wh A^I_{\check a}
A^{I+12}_\mu, \nonumber \\
&& G_{\mu\nu} = G^\ten_{\mu\nu} - G^\ten_{\check a\mu} 
G^\ten_{\check b \nu} \wh
G^{\check a\check b}, \nonumber \\
&& B_{\mu\nu} = B^\ten_{\mu\nu} - 4\wh B_{\check a
\check b} A^{\check a-3}_\mu
A^{\check b-3}_\nu - 2 (A^{\check a-3}_\mu A^{
\check a+3}_\nu - A^{\check a-3}_\nu 
A^{\check a+3}_\mu),
\nonumber \\
&& \Phi = \Phi^\ten - {1\over 4} \ln\det \wh G, \nn
&& \wh \psi_{\check a} = \psi^\ten_{\check a}, \quad
\wt\psi_\mu = \psi^\ten_\mu - 2\, \wh \psi_{\check a} A^{
\check a-3}_\mu\, .
\een
Here $\wh G^{\check a
\check b}$ denotes the inverse of the matrix $\wh G_{\check a
\check b}$.
We have not displayed the spinor indices explicitly; it is 
enough to note that under this dimensional reduction the spinor 
representation of the ten dimensional
rotation group splits into a product of a spinor representation
of the four dimensional rotation group and a spinor
representation of the $SO(6)$ R-symmetry group.
We now combine the scalar fields $\wh G_{\check a
\check b}$, $\wh B_{\check a
\check b}$, and
$\wh A_{\check a}^I$ into an  
$O(6,22)$ matrix valued scalar field $M$.
For this we regard $\wh G_{\check a\check b}$, $\wh B_{\check a
\check b}$ and $\wh A^I_{\check a}$ as
$6\times 6$, $6\times 6$, and $6\times 16$ matrices
respectively, $\wh C_{\check a\check b} = {1\over 2} \wh 
A^I_{\check a}
\wh A^I_{\check b}$ as a 
$6\times 6$ matrix, and define $M$ to be the
$28\times 28$ dimensional matrix
\be \label{newa1}
M = \pmatrix{\displaystyle \wh G^{-1} & \wh G^{-1} (\wh B + \wh
C) & \wh G^{-1}\wh A \cr (-\wh B + \wh C) \wh G^{-1} & (\wh G
- \wh B +
\wh C) \wh G^{-1} (\wh G + \wh B + \wh C) & (\wh G -\wh B +\wh
C)\wh G^{-1} \wh A \cr  \wh A^T \wh G^{-1} &  \wh A \wh G^{-1}
(\wh G + \wh B +\wh
C) & I_{16} + \wh A^T \wh G^{-1} \wh A \cr }\, .
\ee
$M$ satisfies
\be \label{1.1}
M L M^T = L, \quad \quad  M^T=M, \quad \quad L =\pmatrix{0 & I_6
& 0  \cr I_6 & 0 & 0 \cr 0 & 0 & -I_{16}},
\ee
where  $I_n$ denotes the $n\times n$ identity matrix.

The effective action that governs the dynamics of the massless
fields in the four dimensional theory is obtained by
substituting the expressions for the ten dimensional fields in
terms of the four dimensional fields in eq.(\ref{new1}), and taking all
field configurations to be independent of the internal coordinates. The
result is
\ben \label{1.3}
S &=& {1\over 2\pi\alpha'} 
\int d^4 x \sqrt{\det G} \, e^{-2\Phi} \bigg[ R_G +4
G^{\mu\nu}
\p_\mu \Phi \p_\nu\Phi -{1\over 12} G^{\mu\mu'} G^{\nu\nu'}
G^{\rho\rho'} H_{\mu\nu\rho} H_{\mu'\nu'\rho'} \nonumber \\
&&\quad\quad  - G^{\mu\mu'} G^{\nu\nu'} F^i_{\mu\nu} (LML)_{ij}
F^j_{\mu'\nu'} + {1\over 8} G^{\mu\nu} Tr (\p_\mu M L \p_\nu
M L) 
 \bigg] + S_f 
\een
where $S_f$ denotes the fermionic terms, 
$R_G$ is the scalar curvature associated with the four
dimensional metric $G_{\mu\nu}$, and
\ben \label{1.5}
F^i_{\mu\nu} &=& \p_\mu A^i_\nu - \p_\nu A^i_\mu \nonumber \\
H_{\mu\nu\rho} &=& (\p_\mu B_{\nu\rho} + 2 A^i_\mu
L_{ij} F^j_{\nu\rho}) + \hbox{cyclic permutations of
$\mu$, $\nu$, $\rho$}\, .
\een
In deriving this result we have taken
$\int d^6 y=(2\pi\sqrt{\alpha'})^6$, 
where $y^m$ ($1\le m\le 6$) denote the coordinates
labeling the six dimensional torus. Note that we have not written
down the fermionic terms explicitly. Instead of writing the four
dimensional action involving the fermions we shall find it
more convenient to evaluate the quadratic term in the fermions
in the black hole background by directly using the ten dimensional
action \refb{new1}.

For our analysis it will be convenient to use a new set of field variables
which are related to the ones described above by a rotation in the
internal space. We define
\be \label{eudef}
U = \pmatrix{I_6/\sqrt 2 & I_6/\sqrt 2 & \cr -I_6/\sqrt 2 & I_6
/\sqrt 2 & \cr
& & I_{16}}\, ,
\ee
and
\be \label{efielddef}
\bar A^i_\mu = U_{ij} A^j_\mu, \qquad \bar M = UMU^{-1},
\qquad \bar L = ULU^{-1} = \pmatrix{I_6 & \cr & -I_{22}},
\qquad \bar F^i_{\mu\nu} = U_{ij} F^j_{\mu\nu}\, ,
\ee
so that
\be \label{embarlbar}
\bar M \bar L \bar M^T = \bar L, \quad \bar M^T = \bar M\, .
\ee
In these new variables the action takes the form:
\ben \label{1.3a}
S &=& {1\over 2\pi\alpha'} 
\int d^4 x \sqrt{\det G} \, e^{-2\Phi} \bigg[ R_G +4
G^{\mu\nu}
\p_\mu \Phi \p_\nu\Phi -{1\over 12} G^{\mu\mu'} G^{\nu\nu'}
G^{\rho\rho'} H_{\mu\nu\rho} H_{\mu'\nu'\rho'} \nonumber \\
&&\quad\quad  - G^{\mu\mu'} G^{\nu\nu'} \bar
F^i_{\mu\nu} (\bar L \bar M\bar L)_{ij}
\bar F^j_{\mu'\nu'} + {1\over 8} G^{\mu\nu} Tr (\p_\mu 
\bar M \bar L \p_\nu
\bar M \bar L) \bigg] + S_f\, ,
\een
\be \label{edefnewh}
H_{\mu\nu\rho} = (\p_\mu B_{\nu\rho} + 2 \bar A^i_\mu
\bar L_{ij} \bar F^j_{\nu\rho}) + \hbox{cyclic permutations of
$\mu$, $\nu$, $\rho$},
\ee
Finally we can arrive at a simpler version of the action by 
dualizing the 3-form field via the relation
\be \label{ehdual}
H^{\mu\nu\rho} = - i\, (\sqrt{\det G})^{-1} e^{2\Phi}
\epsilon^{\mu\nu\rho\sigma} \p_\sigma \Psi\, ,
\ee
where $\Psi$ is a scalar field. The new action is then given by
\ben \label{1.3b}
S &=& {1\over 2\pi\alpha'} 
\int d^4 x \sqrt{ \det G} \, e^{-2\Phi} \bigg[ R_G +{4}
G^{\mu\nu}
\p_\mu \Phi \p_\nu\Phi -{1\over 2}\, e^{4\Phi} G^{\mu\nu} 
\p_\nu\Psi \p_\mu\Psi \nonumber \\
&&\quad\quad  - G^{\mu\mu'} G^{\nu\nu'} \bar
F^i_{\mu\nu} (\bar L \bar M\bar L)_{ij}
\bar F^j_{\mu'\nu'} +\Psi \, e^{2\Phi}\, G^{\mu\mu'} G^{\nu\nu'}
\bar
F^i_{\mu\nu} \bar L_{ij}
\wt{\bar F}^j_{\mu'\nu'}\nn
&& 
+ {1\over 8} G^{\mu\nu} Tr (\p_\mu 
\bar M \bar L \p_\nu
\bar M \bar L) \bigg] + S_f\, ,
\een
\be \label{edeffdual}
\wt{\bar F}^{i\mu\nu} \equiv {i\over 2}\,
\left(\sqrt{\det G}\right)^{-1} \, \eps^{\mu\nu\rho\sigma}
\bar F^i_{\rho\sigma}\, .
\ee

\subsectiono{The quadratic action for the fluctuations around the
attractor geometry}

We shall consider black holes carrying (electric,magnetic)
charge vectors\footnote{While this is a very specific choice
of the charges, and pair of charges $(\bar Q, \bar P)$, satisfying
$\bar Q^2\equiv \bar Q^T \bar L\bar Q>0$, $\bar P^2>0$
and $\bar Q^2 \bar P^2 > (\bar Q\cdot \bar P)^2$, can be
brought to this form with the help of a continuous
$SL(2,R)\times O(6,22)$ transformation which is a
symmetry of the supergravity equations of motion.
Thus the final result of our analysis holds for any
$(\bar Q,\bar P)$ satisfying $\bar Q^2>0$, $\bar P^2>0$,
$\bar Q^2 \bar P^2 > (\bar Q\cdot \bar P)^2$, --
conditions under which a supersymmetric black hole
solution exists.}
\be \label{ech1}
\bar Q=\pmatrix{Q_0\cr 0\cr \cdot \cr \cdot \cr \cdot \cr 0}, \qquad
\bar P=\pmatrix{0\cr P_0\cr \cdot \cr \cdot \cr \cdot \cr 0}\, .
\ee
Then in an appropriate 
normalization convention the near horizon geometry is given 
by\cite{0708.1270}:
\ben \label{ehor}
&& {ds^2  = a^2 \left(d\eta^2 + \sinh^2\eta \, d\theta^2
\right)   
+ a^2 (d\psi^2+\sin^2\psi\, d\phi^2) \, , \quad
a^2 ={\alpha'\over 8} P_0^2\, }, \quad  
 e^{-2\Phi} 
 ={Q_0\over P_0}, \quad  \bar M=I_{28}, \nonumber \\
&&  {\bar F^{i}_{\eta\theta} = -i{\sqrt{\alpha'}\over 4}\sinh\eta\, 
  {P_0\over Q_0}\, \bar Q_i,} \quad  
 \bar F^{i}_{\psi\phi}={\sqrt{\alpha'} \over 4}
 \bar  P_i\, \sin\psi\, , \quad H_{\mu\nu\rho}=0\, .
 \een
We shall make the choice of vierbeins given in \refb{evier1} and denote
the indices labelling the coordinates of $S^2$ by $\alpha, \beta,\cdots$,
the indices labelling the coordinates of $AdS_2$ by $m,n,\cdots$ and the
indices labelling all the four cordinates by $\mu,\nu,\cdots$.

We shall study quadratic fluctuations of various fields around
the background \refb{ehor}. 
Let us denote the background values of various fields 
given in \refb{ehor} by the superscript $(0)$.
We parametrize
the bosonic fluctuations as follows:
\be \label{efluc}
\Phi = \Phi^{(0)} + {1\over 2}\, \chi_2,  \quad 
G_{\mu\nu} = e^{\chi_2} \left\{G^{(0)}_{\mu\nu} 
+ h_{\mu\nu}\right\},
\quad \Psi = {Q_0\over P_0}\, \chi_1, \quad
\bar A^{i}_\mu 
= \bar A^{i(0)}_\mu + {1\over 2}\, \AAA^{(i)}_\mu\, .
\ee
Special care is taken to parametrize $\bar M$ since it is a 
constrained field. A parametrization satisfying \refb{embarlbar}
to quadratic order in the fluctuations
is as follows:
\be \label{emexp}
\bar M_{ar} = \bar M_{ra} =
\sqrt 2\, \phi_{ar}, \quad \bar M_{ab} = \delta_{ab} + 
\phi_{ar} \phi_{br}, \quad \bar M_{rs} =
\delta_{rs} + \phi_{ar} \phi_{as},
\quad 1\le a,b\le 6, \quad 7\le r,s\le 28\, .
\ee
We also need to add to the action
the gauge fixing term: 
\ben \label{egaugefix}
&& {1\over 2\pi\alpha'}\, {Q_0\over P_0}\, 
\int d^4 x \, \sqrt{\det G^{(0)}}\, \LL_{gf} \, , \nn
\LL_{gf} &=& -{1\over 2} 
g^{\rho\sigma}\,
\left(D^\mu h_{\mu\rho} -{1\over 2} D_\rho \, h^\mu_{~\mu}\right)
\left( D^\nu \, h_{\nu\sigma} -{1\over 2} 
D_\sigma h^\nu_{~\nu}\right)
 - {1\over 2} D^\mu \AAA^{(i)}_\mu D^\nu \AAA^{(i)}_\nu\, ,
\een 
where 
all covariant derivatives and raising of lowering of indices
are computed with the background $AdS_2\times S^2$ metric.
Substituting \refb{efluc}, \refb{emexp} into 
\refb{1.3b}, adding to it \refb{egaugefix}
and expanding it to quadratic order in the
fluctuations we get
\be \label{eactfl}
S   = S^{(0)} + {1\over 2\pi\alpha'}\, {Q_0\over P_0}\, 
\int d^4 x \, \sqrt{\det G^{(0)}} \, \left[
\LL_{standard} + \LL_{flux} 
\right] +S_f\, .
\ee
Here $\LL_{standard}$ is the standard gauge fixed
action for various free
quantum fields in the $AdS_2\times S^2$ background metric:
\ben \label{estandard}
\LL_{standard} &=&  -{1\over 4} h_{\mu\nu} 
\left(\wt \Delta
h\right)^{\mu\nu}+{1\over 2}\, \chi_1\, \square\, \chi_1
+ {1\over 2}\, \chi_2\, \square\, \chi_2+ {1\over 2} \sum_{a=1}^{6}
\AAA_\mu^{(a)} (G^{(0)\mu\nu}\square  - R^{\mu\nu}) 
\AAA_\nu^{(a)}\nn
&& + {1\over 2} \sum_{r=7}^{28}
\AAA_\mu^{(r)} (G^{(0)\mu\nu}\square - R^{\mu\nu}) 
\AAA_\nu^{(r)}
 + {1\over 2} \, \sum_{a=1}^6 \sum_{r=7}^{28}
\phi_{ar} \, \square \, \phi_{ar}\, ,
\een
where,
\ben \label{e35}
\left(\wt\Delta h\right)_{\mu\nu} 
&=& -\square h_{\mu\nu} - R_{\mu\tau} h^\tau_{~\nu}
- R_{\nu\tau} h_\mu^{~\tau} - 2 R_{\mu\rho\nu\tau} h^{\rho\tau}
+{1\over 2} \, G^{(0)}_{\mu\nu} \, 
G^{(0)\rho\sigma} \, \square\, h_{\rho\sigma}
\nn && + R\, h_{\mu\nu} 
+ \left(g_{\mu\nu} R^{\rho\sigma} + R_{\mu\nu}
g^{\rho\sigma}\right) h_{\rho\sigma} -{1\over 2}\, R\, 
g_{\mu\nu} \, g^{\rho\sigma}\, h_{\rho\sigma}
\, .
\een
Here  all indices are 
lowered and raised by the background metric $G^{(0)}_{\mu\nu}$ 
and its
inverse, and $R_{\mu\nu\rho\sigma}$, $R_{\mu\nu}$ and $R$
are calculated with the metric $G^{(0)}_{\mu\nu}$.
$\LL_{flux} $ denotes the extra terms due to
background flux:
\ben \label{eflux}
\LL_{flux} 
&=&  2 \, a^{-1} \, \phi_{2r} \, \vareps^{\gamma\beta} 
\p_\gamma \AAA^{(r)}_\beta 
- a^{-2} \, \phi_{2r} \, \phi_{2r} - 2 i\,
\, a^{-1} \, \phi_{1r} \, \vareps^{mn} 
\p_m \AAA^{(r)}_n + a^{-2} \, \phi_{1r} \, \phi_{1r} \nn
&& + {1\over 2} \, a^{-2}\, \left(h^{mn} h_{mn}
- h^{\alpha\beta} h_{\alpha\beta}
+ 2\, \chi_2\, (  h^m_{~m}
-  h^\alpha_{~\alpha})\right) + {\sqrt 2\over  a}
\left[  i \vareps^{mn}\, f^{(1)}_{\alpha m} h^\alpha_{~n}
+  \vareps^{\alpha\beta}\, f^{(2)}_{\alpha m} h_\beta^{~m}
\right]\nn
&& + {1\over 2\sqrt 2 \, a}\, \left[ i \vareps^{mn} f^{(1)}_{mn} 
\left( -2\chi_2 + h^\gamma_{~\gamma} - h^p_{~p}\right)
- \vareps^{\alpha\beta} f^{(2)}_{\alpha\beta}\, 
\left(  -2\chi_2 + h^p_{~p} - h^\gamma_{~\gamma}\right)
\right] \nn
&& +{1\over a\sqrt 2} \, \chi_1 \, \left( i\ve^{mn} f^{(2)}_{mn}
+ \ve^{\alpha\beta} f^{(1)}_{\alpha\beta}\right)
\nn
f^{(i)}_{\mu\nu} &\equiv& \p_\mu \AAA^{(i)}_\nu
-\p_\nu \AAA^{(i)}_\mu\, .
\een
Here  $\vareps^{\alpha\beta}$ and $\vareps^{mn}$
are the invariant antisymmetric tensors on $S^2$ and $AdS_2$
respectively, computed with the background metric
$G^{(0)}_{\mu\nu}$:
\be \label{edefve}
\ve_{\psi\phi} = a^{2}\, \sin\psi, \qquad \ve_{\eta\theta}
= a^{2} \, \sinh\eta\, .
\ee
$\LL_{standard}$ is what was used in
\S\ref{sother} for computing the heat kernel of various fields;
so our main goal will be to compute the effect of $\LL_{flux}$
on the heat kernels. 

Note that both $\LL_{standard}$ and $\LL_{flux}$ have an 
$SO(22)$ symmetry acting on the index $r$. This is a remnant of
the $SO(6,22)$ continuous duality
symmetry of the original supergravity
action \refb{1.3a} of which an $SO(4,22)$ subgroup survives in the
background \refb{ehor}. 
The gauge fixing term \refb{egaugefix} also respects this symmetry.
The various fluctuations transform either
as a singlet or a vector of $SO(22)$. We shall call the singlets
of $SO(22)$ fields in the gravity multiplet and the vectors 
of $SO(22)$ fields in the matter multiplet. Clearly we can analyze
separately the heat kernels from fields in the matter multiplet and
the gravity multiplet since they will not mix at the quadratic level.
In this paper we shall focus on the contribution from the matter
multiplets only.

To this action we must also add the action for the ghosts associated
with diffeomorphism and U(1) gauge invariances.
Let us denote by $b_\mu$ and $c_\mu$ the ghosts associated
with diffeomorphism invariance, by $b^{(i)}$ and $c^{(i)}$ the ghosts
associated with the U(1) gauge invariances, and by
\be \label{egfterm}
\FF_\mu \equiv D^\rho h_{\mu\rho} -{1\over 2} D_\mu h^\rho_{~\rho},
\qquad 
\FF^{(i)} \equiv D^\rho \AAA^{(i)}_\rho
\ee
the gauge fixing terms for the diffeomorphism and U(1) gauge invariances.
Then by standard rules the ghost lagrangian density will be given by
\be \label{eghostaction}
b^\mu \delta\FF_\mu + 
+ b^{(i)} \delta \FF^{(i)} \, ,
\ee
where $\delta$ denotes the variation under a
diffeomorphism transformation with
parameter $c_\nu$ and gauge transformations
with parameters $c^{(i)}$.
In the attractor geometry given in \refb{ehor} we have
\be \label{etrslaw}
\delta h_{\mu\nu} = D_\mu c_\nu + D_\nu c_\mu+\cdots, \qquad
\delta \AAA^{(i)}_\mu = D_\mu c^{(i)}
- 2\, \bar F^i_{\mu\nu} c^\nu + 2\, 
D_\mu (\bar A^{(i)}_\rho c^\rho)
+\cdots\, ,
\ee
where $D_\mu$ denotes covariant derivative computed using
the background metric of the attractor geometry, $\bar A^{(i)}_\mu$
is the background gauge field (note the normalization factor of
2 between the background and fluctuations given in \refb{efluc})  and
$\cdots$ denotes terms higher order in the fluctuations. Using
\refb{egfterm}-\refb{etrslaw}, and the fact that the background
geometry satisfies $D^\mu \bar F^i_{\mu\nu}=0$, 
 we see that
the quadratic part of the ghost action is given by
\be \label{eqghost}
b^\mu \left(g_{\mu\nu}\square + R_{\mu\nu}\right) 
c^\nu + b^{(i)}\square c^{(i)}
- 2 \, b^{(i)} \bar F^i_{\mu\nu} \, D^\mu c^\nu + 2\, b^{(i)}
\, \square \, (\bar A^i_\rho c^\rho)
\ee
The last term can be removed by a field redefinition 
$c^{(i)}\to c^{(i)}-2\, \bar A^i_\rho c^\rho$. This yields a
simpler version of the ghost lagrangian density:
\be \label{emghost}
\LL_{ghost} \propto 
\left[b^\mu \left(g_{\mu\nu}\square + R_{\mu\nu}\right) 
 c^\nu + b^{(i)}\square c^{(i)}
- 2 \, b^{(i)} \bar F^i_{\mu\nu} \, D^\mu c^\nu\right]\, .
\ee
Note that the ghosts $(b^{(i)}, c^{(i)})$ for $i=7,\cdots 28$
are $SO(22)$ vectors and the rest of the ghosts are $SO(22)$
singlets. As in the case of matter fields, we shall analyze the
contribution from $SO(22)$ vector ghosts only.

Finally let us turn to the fermionic terms in the action. In order
to simplify the structure of the action we need to carry out a set
of field redefinitions. We define
\ben \label{eferdef}
\psi_\mu &=& \wt\psi_\mu +{1\over 2} \Gamma_\mu \, 
\Gamma^{\check a}
\wh \psi_{\check a}\, , \nn
\lambda &=& {1\over 2}\left(
\Lambda + \sqrt 2 \, \Gamma^{\check a} \wh\psi_{\check a}
 \right)\, , \nn
\vp_{{\check a}+3} &=& \wh \psi_{\check a} 
- {1\over 2\sqrt 2} \, \Gamma_{\check a}\Lambda , \nn
\vp_{I+12} &=& \Sigma^I \, , \qquad
 0\le \mu, \nu \le 3, \quad 4\le {\check a}, {\check b}\le 9, \quad
1\le I\le 16\, ,
\een
where $\wh\psi_{\check a}$ and $\wt\psi_\mu$ have been defined
in \refb{1.2}.
Then the quadratic terms in the fermionic action,
evaluated in the background \refb{ehor},  takes the form
\be \label{efaction}
S_f = {1\over 2\pi\alpha'}\, {Q_0\over P_0}\, \int d^4 x \, 
\sqrt{\det G^{(0)}}\, \LL_{f}\, ,
\ee
where
\ben \label{efermi}
\LL_f &=& -{1\over 2} \Bigg[ \sum_{r=7}^{28} 
\bar \vp_r \left\{\Gamma^\mu D_\mu +
{1\over 2\sqrt 2} \, \Gamma^{\rho\sigma}  
 \left(\bar F^1_{\rho\sigma} \Gamma^4 
+\bar F^2_{\rho\sigma} 
\Gamma^5\right)
\right\} \vp_r 
\nn  
&& + \bar \psi_\mu \Gamma^{\mu\nu\rho} D_\nu
\psi_\rho + \bar \lambda \Gamma^\mu D_\mu \lambda \nn
&& 
+  {1\over 4\sqrt 2}\, \bar\psi_\mu 
\left[-\Gamma^{\mu\nu\rho\sigma}
+ 2 g^{\mu\sigma} g^{\nu\rho} + 2 \Gamma^{\mu\rho\nu}
\Gamma^\sigma + \Gamma^{\mu\nu}\Gamma^{\rho\sigma}
\right] \left(\bar F^1_{\rho\sigma} \Gamma^4 
+\bar F^2_{\rho\sigma} 
\Gamma^5\right)  \psi_\nu \nn &&
+ {1\over 4} \left[ \bar\psi_\mu \Gamma^{\rho\sigma}
\Gamma^\mu \left(\bar F^1_{\rho\sigma} \Gamma^4 
+\bar F^2_{\rho\sigma} 
\Gamma^5\right) \lambda
-\bar\lambda \left(\bar F^1_{\rho\sigma} \Gamma^4 
+\bar F^2_{\rho\sigma} 
\Gamma^5\right) \Gamma^\mu
\Gamma^{\rho\sigma} \psi_\mu
\right] 
\Bigg]\, .
\een
To this we add the gauge fixing term
\be \label{efgf}
{1\over 2\pi\alpha'} \, {Q_0\over P_0}\, \int d^4 x \, 
\sqrt{\det G^{(0)}}\, \LL'_{gf}\, ,
\ee
where
\be \label{elagfgf}
\LL'_{gf} = {1\over 4} 
\bar \psi_\mu \Gamma^\mu \Gamma^\nu 
D_\nu\Gamma^\rho
 \psi_\rho\, .
\ee

The structure of the ghost action can be determined as follows.
Since the gauge fixing term is $\Gamma^\mu\psi_\mu$, the
lagrangian density for the spinor valued bosonic ghost fields
$\tilde b$, $\tilde c$ is proportional to $\bar {\tilde b}
\Gamma^\mu \delta\psi_\mu$, where $\delta\psi_\mu$ is the
variation of $\psi_\mu$ under the supersymmetry transformation
with parameter $\tilde c$. Using the supersymmetry transformation
laws of the ten dimensional fields given in \refb{esusygrav}, 
and the relations
between the ten and the four dimensional fields given in
\refb{eferdef}, we find that 
in the near horizon background \refb{ehor},
\ben \label{epsitrs}
\delta \psi_\mu &=& D_\mu \tilde c +{1\over 4\sqrt 2}\,
 \left( 4 \delta_\mu^{\rho} \Gamma^\sigma
- \Gamma_\mu \Gamma^{\rho\sigma}\right) 
\left(
\bar F^{(1)}_{\rho\sigma} \Gamma^{4}
+ \bar F^{(2)}_{\rho\sigma} \Gamma^{5}\right)
\tilde c + \cdots \nn
\delta\lambda &=& -{1\over 4}\, \Gamma^{\rho\sigma}\,
\left(
\bar F^{(1)}_{\rho\sigma} \Gamma^{4}
+ \bar F^{(2)}_{\rho\sigma} \Gamma^{5}\right)
\tilde c\, .
\een
where $\cdots$ denotes terms which vanish in the background
\refb{ehor}. Thus 
\be \label{epsitr2}
\Gamma^\mu\delta \psi_\mu = \Gamma^\mu D_\mu \tilde c\, ,
\ee
and
the ghost lagrangian density is proportional to
\be \label{efgho}
\LL_{f;ghost}\propto \bar{\tilde b} \, \Gamma^\mu D_\mu \tilde c
\, .
\ee
Quantization of the gravitino also requires the introduction
of a third spin 1/2 bosonic 
ghost field. This comes from the
special nature of the gauge fixing term given in \refb{elagfgf}; to get this
term we first insert into the path integral
the gauge fixing term 
$\delta(\Gamma^\mu \psi_\mu - \xi(x))$
for some
arbitrary space-time dependent spinor $f(x)$; 
and then average over all $\xi(x)$ with a weight factor of
$\exp(-\bar \xi \not \hskip -4pt D \xi)$. The integration over
$\xi$ introduces an extra factor of $\det \not \hskip -4pt D$ which
needs to be canceled by an additional spin half bosonic ghost
with the standard kinetic operator
proportional to $\not \hskip -4pt D$. Denoting the new ghost field
by $\tilde e$ we get the additional ghost action to be
\be \label{eaddghost}
\LL_{f;ghost}'\propto \bar{\tilde e} \, \Gamma^\mu D_\mu \tilde e
\, .
\ee

In the fermionic sector only the fields $\vp_r,\bar\vp_r$ are
$SO(22)$ vectors. Rest of the fields including all the ghosts
are $SO(22)$ singlets. As before we shall analyze the
contribution to the one loop effective action
from $SO(22)$ vector fields only.

\renewcommand{\theequation}{\thesection.\arabic{equation}}

\sectiono{Eigenvalues, heat kernel and  one loop effective
action in the matter sector} \label{sfive}

In this section we 
shall compute the eigenvalues and eigenfunctions
of the kinetic operator in the matter sector and use it to calculate the
logarithmic correction to the extremal black hole entropy.
The $SO(22)$ symmetry
guarantees that at the quadratic level
there is no mixing between fields carrying different
$r$ values, so we can analyze one $r$ value at a time.

We first focus on the bosonic fields. From the structure of the
action we see that the fields $\phi_{ar}$ for $3\le a\le 6$ do not
enter $\LL_{flux}$; so their heat kernel is given by the standard
heat kernel of scalar fields computed with $\LL_{standard}
+ \LL_{gf}$. The field $\phi_{2r}$ mixes with the
component of $\AAA^{(r)}$ along $S^2$ and the field
$\phi_{1r}$ mixes with the component of $\AAA^{(r)}$
along $AdS_2$. Thus we can separately analyze these two
cases.  This reduces the problem to that of a mixing between a
single scalar and a vector field.

First we shall consider the mixing between $\phi_{2r}$
and the component of $\AAA^{(r)}$ along $S^2$.
To avoid proliferation of indices we drop the 
$SO(22)$ and $SO(6)$ indices on
the fields, define $g_{\alpha\beta}\equiv G^{(0)}_{\alpha\beta}$ 
and express
the relevent quadratic term in the action
as
\be \label{emix1}
-{1\over 2} \int\sqrt{\det g}\, 
\left[ \pmatrix{\phi & \AAA_\alpha} \pmatrix{
-\square + 2 \, a^{-2} & -2 \, a^{-1} \vareps^{\gamma\beta} 
D_\gamma\cr
-2 \, a^{-1}\vareps^{\alpha\gamma} D_\gamma  &
- g^{\alpha\beta}\square +R^{\alpha\beta}
+ D^\alpha D^\beta } \pmatrix{
\phi\cr \AAA_\beta} - \AAA_\alpha D^\alpha D^\beta 
\AAA_\beta\right]\, ,
\ee
up to an overall multiplicative factor.
The last term is the gauge 
fixing term.
In order to construct the heat kernel of the combined system of the
scalar and the gauge fields we need to find the eigenstates of the
kinetic
operator appearing in \refb{emix1}. 
For this we first decompose the gauge field as
\be \label{emix2}
\AAA_\alpha = D_\alpha \psi + \vareps_\alpha^{~\beta}
D_\beta \chi\, ,
\ee
where $\psi$ and $\chi$ are scalars on $S^2$. Substituting this
into \refb{emix1} we get
\ben \label{emix3}
&& -{1\over 2} \int\sqrt{\det g}\,
\bigg[ \pmatrix{\phi & 
\vareps_\alpha^{~\alpha\prime}D_{\alpha\prime} \chi} \pmatrix{
-\square + 2 \, a^{-2} & -2 \, a^{-1} \vareps^{\gamma\beta} 
D_\gamma\cr
-2 \, a^{-1}\vareps^{\alpha\gamma} D_\gamma  &
-g^{\alpha\beta} \square +R^{\alpha\beta}
+ D^\alpha D^\beta } \pmatrix{
\phi\cr \vareps_\beta^{~\beta\prime}D_{\beta\prime} \chi} 
\nn && 
\qquad - D_\alpha\psi D^\alpha D^\beta D_\beta\psi\bigg]\, .
\een
Note that the part of $\AAA_\alpha$ involving $\psi$ does not
contribute to the first term and the part of $\AAA_\alpha$
involving $\chi$ does not contribute to the second term.
We now decompose $\phi$, $\psi$ and $\chi$ as
\be \label{emix4}
\phi = \sum_n a_n f_n(x), \qquad
\chi = \sum_n {1\over \sqrt \kappa_n} b_n f_n(x), \qquad
\psi = \sum_n {1\over \sqrt \kappa_n} c_n f_n(x),
\ee
where $\{f_n\}$ is an orthonormal basis of 
eigenfunctions of the scalar $-\square$ operator
with eigenvalues $\{\kappa_n\}$. Substituting this into
\refb{emix3} we get
\be \label{emix5}
-{1\over 2} \sum_n \left[\pmatrix{a_n & b_n} \pmatrix{
\kappa_n + 2 a^{-2}& -2 a^{-1} \sqrt{\kappa_n} \cr 
-2 a^{-1} \sqrt{\kappa_n} & \kappa_n}\pmatrix{a_n\cr b_n}
+ \kappa_n c_n c_n\right]\, .
\ee
This shows that the eigenvalues of the modes labelled by $c_n$ are
not affected by the mixing with the scalars. On the other hand
the eigenvalues of the modes labelled by $a_n$, $b_n$ change from
$\kappa_n$ to 
\be\label{ev1}
\kappa_n + a^{-2}  \pm a^{-1} \sqrt{4\kappa_n+a^{-2}}\, ,
\ee
for $\kappa_n>0$.
For $\kappa_n=l(l+1) a^{-2}$ with $l>0$
this gives the eigenvalues
\be \label{ev2}
(l-1)la^{-2}, \qquad (l+1)(l+2)a^{-2}\, .
\ee
Thus for each pair of modes with $l>0$, one has its $l$ value
shifted by $+1$ and one has its $l$ value shifted by $-1$.
Finally for $l=0$ there are no modes from $b_n$, and the eigenvalue
of the mode $a_n$ shifts from 0 in the absence of flux to $2a^{-2}$.
This effectively causes a shift of the $l=0$ eigenvalue to
$l=1$ eigenvalue.  
Thus the net additional 
contribution to the trace of the
heat kernel on $S^2$
from the scalar and the gauge
field is given by
\ben \label{eshift1}
\delta K^{v+s}_{S^2}&=&{1\over 4\pi a^2}
\left[\sum_{l=0}^\infty (2l+1) 
\left\{ e^{-s (l+1) (l+2)/a^2} -e^{-sl(l+1)/a^2}  \right\} \right.
\nn && \left.
+\sum_{l=1}^\infty (2l+1) \left\{ e^{-s (l-1) l/a^2} -e^{-sl(l+1)/a^2} 
\right\}
\right]
\, ,
\een
where the $l=0$ term in the first sum takes into account the shift
of the scalar mode with $l=0$ to $l=1$. We now break this as
a sum of four different sums and 
shift $l\to l\mp 1$ in the first and the third terms.
This gives 
\ben \label{eshift2}
\delta K^{v+s}_{S^2}&=& {1\over 4\pi a^2}
\left[\sum_{l=1}^\infty (2l-1) \, e^{-sl(l+1)/a^2}
-\sum_{l=0}^\infty (2l+1) \, e^{-sl(l+1)/a^2}\right. \nn
&& \qquad \qquad \left.
+ \sum_{l=0}^\infty (2l+3) \, e^{-sl(l+1)/a^2}
- \sum_{l=1}^\infty (2l+1) \, e^{-sl(l+1)/a^2}\right] \nn
&=& {1\over 2\pi a^2}\, .
\een
Thus the net contribution to the heat kernel from a vector on
$S^2$ and the scalar on $S^2$ with which the vector mixes
is given by
\be \label{eshift3}
K^{v+s}_{S^2}= 
K^v_{S^2} + K^s_{S^2} + {1\over 2\pi a^2} 
= 3 \, K^s_{S^2}\, ,
\ee
where in the last step we have used
\refb{e21a}.

A similar analysis can be done for the mixing between 
$\phi_{1r}$ and the component of the vector
field $\AAA^{(r)}$ along $AdS_2$. The main difference between the
$S^2$ and the $AdS_2$ case is that for $AdS_2$ the mixing term
\refb{emix1} is replaced by
\be \label{emix1a}
-{1\over 2} \int\sqrt{\det g}\,
\left[ \pmatrix{\phi & \AAA_m} \pmatrix{
-\square - 2 \, a^{-2} & 2 \, i\, a^{-1} \vareps^{pn} 
D_p\cr
2 \, i\, a^{-1}\vareps^{mp} D_p  &
- g^{mn}\, \square +R^{mn}
+ D^m D^n} \pmatrix{
\phi\cr \AAA_n} - \AAA_m D^m D^n 
\AAA_n\right]\, .
\ee
One can now analyze its effect on the eigenvalues of the kinetic
operator exactly as in the case of $S^2$. The final outcome of this
analysis is that 
if we denote  the eigenvalue of
$(d\delta + \delta d)$ on a vector field of the form
$\varepsilon_n^{~p} D_p\chi$ or a scalar field on $AdS_2$
by $\kappa\equiv \left(\lambda^2 +{1\over 4}\right)/a^2$, 
then acting on fields $\phi$ and $\AAA_n$ carrying this eigenvalue
the kinetic operator coming from the first term takes the form:
\be \label{ekin2}
\pmatrix{\kappa -2 a^{-2}  & 2 i \sqrt{\kappa} a^{-1}\cr
2 i\sqrt{\kappa} a^{-1} & \kappa}\, .  
\ee 
Although this matrix is complex, it is diagonalizable with a 
(complex) orthogonal matrix
and gives eigenvalues 
\be \label{eigennew}
a^{-2} \left[ (\lambda\pm i)^2 +{1\over 4} \right]\, .
\ee
Thus
the mixing between the scalar and the vector on $AdS_2$ shifts the
parameter $\lambda$ by $i$ for one set of states and $-i$ for
another set of states. The kinetic operator in the second
term of \refb{emix1a} continues to have the eigenvalue
$\left(\lambda^2 +{1\over 4}\right)/a^2$ on fields of the form
$D_n\psi$. As a result the
net change in the heat kernel of the scalar and the vector on $AdS_2$
is given by\footnote{Note that as $\lambda\to 0$ the integrand
grows as $\exp(3\bar s/4)$. Thus the eigenvalues of the kinetic operator
are negative and the path integral is not well defined,
reflected in the fact that the integration over $s$ will diverge for
large $s$ if we try to carry out
the integration over $s$ first for a fixed $\lambda$.
Physically this divergence is a 
consequence of the imaginary background electric field in the
euclidean
$AdS_2$ space. However as
we shall see, if we carry out the integration over $\lambda$ first then
there is a cancelation and the exponentially divergent contribution
in the large $s$ limit is removed. This procedure is consistent with
the rules for computing loop amplitudes in string theory, where the
integration over the modular parameter (the analog of $s$) is carried
out at the end, after we have integrated / summed over the
eigenvalues of the kinetic operator. Presumably at the level of
the path integral this corresponds to deforming the path integration
contour to the complex configuration space where the path
integral is well defined, as {\it e.g.} in \cite{1001.2933}.
Further justification of this procedure can be found in appendix
\ref{sa}.
\label{ff}}
\be \label{eshift4}
\delta K^{v+s}_{AdS_2}=
{1\over 2\pi a^2} \exp[-\bar s/4]
\, \int_0^\infty \, d\lambda \, \lambda \, 
\tanh(\pi \lambda) \, \left[ e^{-\bar s(\lambda+i)^2}
+ e^{-\bar s(\lambda-i)^2} - 2 \, e^{-\bar s \lambda^2}\right]\, .
\ee
We now shift the integration variable $\lambda\to \lambda\mp i$ in the
first two terms and express this as
\ben \label{eshift5}
\delta K^{v+s}_{AdS_2} &=&
{1\over 2\pi a^2} \exp[-\bar s/4]
\left[ \int_i^{i+\infty} \, d\lambda \, (\lambda-i) \, 
\tanh(\pi \lambda-i\pi) e^{-\bar s \lambda^2} \right. \nn
&& \left.
+  \int_{-i}^{-i+\infty} \, d\lambda \, (\lambda+i) \, 
\tanh(\pi \lambda+i\pi) e^{-\bar s \lambda^2}
 - 2 \int_0^\infty \, d\lambda \, \lambda \, 
\tanh(\pi \lambda) \, e^{-\bar s \lambda^2}
\right] \, . \nn
\een
Using $\tanh(x\pm i\pi)=\tanh x$ and the fact that the integrands
in \refb{eshift5} do not have any poles for $Re(\lambda)>0$, we can
deform the integration contour in the first integral as a contour
from $i$ to 0 lying in the $Re(\lambda)>0$ region and a contour
from 0 to $\infty$ along the real axis. 
Similarly the integration contour in the second
integral can be deformed to a 
contour
from $-i$ to 0 lying
in the $Re(\lambda)>0$ region and a contour
from 0 to $\infty$ along the real axis. 
The total contribution from the contours from 0 to $\infty$ cancel
the last term in \refb{eshift5}, and we get
\be \label{eshift6}
\delta K^{v+s}_{AdS_2} =
{1\over 2\pi a^2} \exp[-\bar s/4] \, \left[
\int_i^{0(+)} \, d\lambda \, (\lambda-i) \, 
\tanh(\pi \lambda) e^{-\bar s \lambda^2}
+ \int_{-i}^{0(+)} \, d\lambda \, (\lambda+i) \, 
\tanh(\pi \lambda) e^{-\bar s \lambda^2}\right]\, ,
\ee
where the superscript $(+)$ denotes that we are integrating along 
a contour in the $Re(\lambda)>0$ region. Making a change
of variables $\lambda \to -\lambda$ in the second term we get
\be \label{eshift7}
\delta K^{v+s}_{AdS_2} =
{1\over 2\pi a^2} \exp[-\bar s/4] \, \left[
\int_i^{0(+)} \, d\lambda \, (\lambda-i) \, 
\tanh(\pi \lambda) e^{-\bar s \lambda^2}
- \int_{i}^{0(-)} \, d\lambda \, (\lambda-i) \, 
\tanh(\pi \lambda) e^{-\bar s \lambda^2}\right]\, .
\ee
We now have a closed clockwise contour. The result of this contour
integral can be easily evaluated in terms of the residue at the pole at
$\lambda = i/2$, and we get
\be \label{eshift7a}
\delta K^{v+s}_{AdS_2} =
-{1\over 2\pi a^2}\, .
\ee
Thus the net contribution to the trace of the
heat kernel from a vector of
$AdS_2$ and the scalar on $AdS_2$ with which the vector mixes
is given by
\be \label{eshift3a}
K^{v+s}_{AdS_2}= 
K^v_{AdS_2} + K^s_{AdS_2} - {1\over 2\pi a^2} 
= 3 \, K^s_{AdS_2}\, ,
\ee
where in the last step we have used \refb{e21b}.

We can now
use these results to compute the net contribution to the
heat kernel from the bosonic fields of a matter multiplet on
$AdS_2\times S^2$. First of all there are four scalars which do
not mix with the vector; their contribution will be given by
$4 K^s_{AdS_2}K^s_{S^2}$. Then we have a vector of $S^2$
which mixes with one of the remaining scalars, giving a contribution
$K^{v+s}_{S^2} K^s_{AdS_2}$ with $K^{v+s}_{S^2}$ given
in \refb{eshift3}. Next we have a vector along $AdS_2$ that mixes
with the remaining scalar and gives a contribution $K^s_{S^2}
K^{v+s}_{AdS_2}$ with $K^{v+s}_{AdS_2}$ given in
\refb{eshift3a}. 
Finally we have a pair of ghosts whose contribution 
$-2K^s_{AdS_2} K^s_{S^2}$ needs to be added.
Thus the net contribution from the six scalars and
one vector of the matter multiplet is given by
\be \label{enetv}
K^{v+6s}_{AdS_2\times S^2}(0;s)
= 4 K^s_{AdS_2}K^s_{S^2} + K^s_{AdS_2}
K^{v+s}_{S^2} + K^s_{S^2}
K^{v+s}_{AdS_2} - 2 K^s_{AdS_2} K^s_{S^2}
= 8\, K^s_{AdS_2}(0;s) \, K^s_{S^2}(0;s)\, .
\ee
Note that the small $s$
expansion of this quantity (and a similar result for the trace of
the fermionic
heat kernel given in \refb{efer7}) could be computed using the heat
kernel expansion discussed {\it e.g.} in \cite{gilkey}. However
\refb{enetv} and \refb{efer7} 
also has information about the large $s$ behaviour.
This is needed to identify and subtract the zero mode contributions.

Let us now consider the effect of the background 
flux on the fermionic
fields in the matter multiplet. These fields are 
the fields $\vp_r$ appearing in \refb{efermi}, and
transform in the  vector representation of $SO(22)$.
{}From \refb{ehor}, \refb{efermi}, and the representations of the
gamma matrices given in \refb{egamma},
we see 
that the Dirac operator acting on
the fermions takes the form\footnote{Although the original
fermions are chiral -- in the sense that their chirality property under
the space-time Lorentz group $SO(4)$ is correlated with their chirality
under the internal R-symmetry group $SO(6)$ --
in order to compute the eigenvalue
of the Dirac operator we shall ignore the chirality projection and
then take appropriate square root of the determinant. 
Since this 
doubles the number of fermionic
degrees of freedom, the action is not manifestly supersymmetric.
We can avoid this
by appropriately pairing the fermions in
the dimensionally reduced four dimensional theory to
construct Dirac fermions without
using any additional fermionic degrees of freedom.
Thus in this
description we can maintain manifest supersymmetry. This is 
essential if we make use of supersymmetry in evaluating the
path integral; {\it e.g.} using localization 
arguments\cite{0905.2686}.
However for the explicit computation of the one loop
determinant the loss of manifest supersymmetry of the
action will not be a problem.} 
\be \label{end0}
\not \hskip -4pt \DD = \not \hskip -4pt \DD_{S^2}
+  \sigma_3 \, \not \hskip -4pt \DD_{AdS_2}\, ,
\ee
where
\be \label{end1}
\not \hskip -4pt \DD_{S^2} =
\not \hskip -4pt D_{S^2} 
- {i\over 2} \, a^{-1} \, \wh\Gamma^5\, \tau_3, \qquad
\not \hskip -4pt \DD_{AdS_2}=
\not \hskip -4pt D_{AdS_2}
- {1\over 2}  a^{-1} \, \wh\Gamma^4
 \, .
\ee
$\not \hskip -4pt D_{S^2}$ and
$\not \hskip -4pt D_{AdS_2}$ have been
defined in \refb{ed1}, \refb{ed1a}, and
$\wh\Gamma^4$ and $\wh\Gamma^5$ are two of the six
$SO(6)$ gamma matrices satisfying
\be \label{end2}
\{\wh\Gamma^i, \wh\Gamma^j\} = 2\, \delta_{ij}, \qquad
[\wh\Gamma^i, \sigma_a] = 0 = [\wh\Gamma^i, \tau_a],
\qquad 1\le a\le 3, \quad 4\le i,j\le 9\, .
\ee
One can easily check 
that $\not \hskip -4pt \DD_{S^2}$ and 
$\sigma_3\not \hskip -4pt \DD_{AdS_2}$ anticommute.
Hence $\not \hskip -4pt \DD^2 = \not \hskip -4pt \DD_{S^2}^2
+  \not \hskip -4pt \DD_{AdS_2}^2$, the eigenvalues of
$\not \hskip -4pt \DD^2$ are given by the sum of the eigenvalues
of $\not \hskip -4pt \DD_{S^2}^2$ and
$\not \hskip -4pt \DD_{AdS_2}^2$, and the trace of the
heat kernel
of $\not \hskip -4pt \DD$ is given by $-1$ times the 
product of the traces of the heat kernels
of $\not \hskip -4pt \DD_{S^2}$
and $\not \hskip -4pt \DD_{AdS_2}$. Thus we 
first need to find the
eigenvalues of $\not \hskip -4pt \DD_{S^2}$
and $\not \hskip -4pt \DD_{AdS_2}$. 
Since $\wh\Gamma^5\tau_3$ and $\wh\Gamma^4\sigma_3$
each have eigenvalues $\pm 1$ and commute with
$\not \hskip -4pt \DD_{S^2}$
and $\not \hskip -4pt \DD_{AdS_2}$ respectively, it follows
from \refb{end1} that the eigenvalues of $\not \hskip -4pt \DD_{S^2}$
are given by the eigenvalues of 
$\not \hskip -4pt D_{S^2}$ $\pm ia^{-1} / 2$, and the eigenvalues
of   $\not \hskip -4pt \DD_{AdS_2}$
are given by the eigenvalues of 
$\not \hskip -4pt D_{AdS_2}$ $\pm a^{-1} / 2$,
Using this result and eqs.\refb{ed3}, \refb{eadsev}  we see that the eigenvalues of 
$\not \hskip -4pt \DD_{S^2}$ are given by 
$\pm i a^{-1} \left(l+1\pm {1\over 2}\right)$ 
and the eigenvalues
of   $\not \hskip -4pt \DD_{AdS_2}$ are given by 
$\pm i a^{-1} \left(\lambda \pm {i\over 2}\right)$. 
As a result $K^f_{S^2}$ defined in \refb{ed7} 
changes to\footnote{We are giving the result for the heat
kernel per Dirac fermion.}
\be \label{efer1}
K^{f\prime}_{S^2}(0;s) =
-{1\over 4\pi a^2} \, \sum_{l=0}^\infty (2l +2)\,
\left[ e^{-s(l+{3\over 2})^2/a^2} + e^{-s(l+{1\over 2})^2/a^2}
\right]
\, ,
\ee
and 
$K^f_{AdS_2}$ given in
\refb{kfads} is replaced by
\ben \label{efer3}
K^{f\prime}_{AdS_2}(0;s)&=& -{1\over 2\pi a^2}
\int_0^\infty d\lambda \left[ e^{-\bar s(\lambda+{i\over 2})^2}
+ e^{-\bar s(\lambda-{i\over 2})^2}\right]
\, \lambda \, \coth(\pi\lambda)\nn
&=& -{1\over 2\pi a^2} \int_{i/2}^{{i\over 2}+\infty} 
d\lambda \,
e^{-s\lambda^2/a^2}
(\lambda-{i\over 2}) \tanh(\pi\lambda) \nn &&
- {1\over 2\pi a^2} \int_{-i/2}^{-{i\over 2}+\infty} 
d\lambda \,
e^{-s\lambda^2/a^2}
(\lambda+{i\over 2}) \tanh(\pi\lambda),\nn
\een
where in the second step we have shifted 
$\lambda\to \lambda -{i\over 2}$ in the first term and 
$\lambda\to \lambda +{i\over 2}$ in the second term.

Changing $l\to l-1$ in the first term in \refb{efer1} and defining
$\bar s = s/a^2$
we get
\be \label{efer2}
K^{f\prime}_{S^2}(0;s)
= -{1\over 4\pi a^2} \, e^{-\bar s / 4}
\sum_{l=0}^\infty \, e^{-\bar s l (l+1) }
(2l + 2l+2) = -{1\over 2\pi a^2} \, e^{-\bar s / 4}
\sum_{l=0}^\infty \, e^{-\bar s l (l+1) } (2l+1)\, .
\ee
On the other hand in \refb{efer3}
we deform the first integration contour to over the
range $i/2$ to 0 and 0 to $\infty$ and the second integration
contour to over the range $-i/2$ to 0 and 0 to $\infty$. This
gives
\ben \label{efer4}
K^{f\prime}_{AdS_2}(0;s)&=& 
-{1\over 2\pi a^2} \int_{i/2}^{0(+)} d\lambda \,
e^{-s\lambda^2/a^2}
(\lambda-{i\over 2}) \tanh(\pi\lambda) \nn &&
-
{1\over 2\pi a^2} \int_{-i/2}^{0(+)} d\lambda \,
e^{-s\lambda^2/a^2}
(\lambda+{i\over 2}) \tanh(\pi\lambda)\nn &&
- {1\over \pi a^2} \int_{0}^{\infty} d\lambda \,
e^{-s\lambda^2/a^2}\, 
\lambda\, \tanh(\pi\lambda)\, .
\een
As before the superscript $(+)$ denotes that the contour lies in
the $Re(\lambda)>0$ region. Changing $\lambda\to -\lambda$ in
the second integral gives
\ben \label{efer5}
K^{f\prime}_{AdS_2}(0;s)&=& 
-{1\over 2\pi a^2} \int_{i/2}^{0(+)} d\lambda \,
e^{-s\lambda^2/a^2}
(\lambda-{i\over 2}) \tanh(\pi\lambda)  \nn && +
{1\over 2\pi a^2} \int_{i/2}^{0(-)} d\lambda \,
e^{-s\lambda^2/a^2}
(\lambda-{i\over 2}) \tanh(\pi\lambda)\nn &&
- {1\over \pi a^2} \int_{0}^{\infty} d\lambda \,
e^{-s\lambda^2/a^2}\, 
\lambda\, \tanh(\pi\lambda)\, .
\een
We now note that the first and second integrals can be combined
into a singe clockwise contour and the result vanishes since
the
integrand does not have any singularity enclosed by the contour.
Thus we have
\be \label{efer6}
K^{f\prime}_{AdS_2}(0;s) = -{1\over \pi a^2} \int_{0}^{\infty} 
d\lambda \,
e^{-s\lambda^2/a^2}\, 
\lambda\, \tanh(\pi\lambda)\, .
\ee
Combining this with \refb{efer2} we get the net contribution to
the effective heat kernel of a Dirac fermion on $AdS_2\times S^2$
in the presence of background flux:
\ben \label{efer7}
K^{f\prime}(0;s) &=& -K^{f\prime}_{AdS_2}(0;s)
K^{f\prime}_{S^2}(0;s) \nn
&=& -{1\over 2\pi^2 a^4} \, e^{-\bar s / 4}
\sum_{l=0}^\infty e^{-\bar s l (l+1)   } (2l+1)
\, \int_{0}^{\infty} \, d\lambda \, 
e^{-\bar s\lambda^2}\, 
\lambda\, \tanh(\pi\lambda)\nn
&=& -4\, K^s(0;s)\, ,
\een
where $K^s(0;s)=K^s_{S^2}(0;s)K^s_{AdS_2}(0;s)$ 
is the heat kernel of a scalar on $AdS_2\times S^2$
as given in \refb{ecomb1}.
Since a single matter multiplet contains four Weyl fermions, or
equivalently two Dirac fermions, the net contribution to the heat
kernel from the fermions is given by
$-8\, K^s(0;s)$. This exactly cancels the contribution \refb{enetv},
showing that the net contribution to the heat kernel from a matter
multiplet is zero. 

Our result also shows that no subtraction of the type
described in \refb{edefkp} is needed to
regulate the infrared divergences.
Mathematically it is a consequence of an additional $s$
independent constant term
that arose in the expression for $\delta K^{v+s}_{AdS_2} K^s_{S^2}$.
However 
physically this is somewhat surprising given that
the subtraction was needed to 
remove the contribution from the zero modes
of the vector fields on $AdS_2\times S^2$. 
In appendix \ref{sa} we have provided a justification of this
procedure by carefully analyzing the contribution from 
integration over these zero modes.
We also
note that since these zero modes transform non-trivially under
a simultaneous rotation in $AdS_2$ and $S^2$, the argument
of \cite{0905.2686} shows that the 
contribution to the path integral due to
these zero modes will cancel a similar contribution from the
fermion zero modes. Put another way, supersymmetry
allows us add a term to
the action which does not change the result of the path integral
but lifts the zero modes. Thus it appears that the analytic
continuation procedure we have adopted, namely doing the
$\lambda$ integral first and then the $s$ integral, automatically
accounts for this cancelation. This clearly deserves further
study.

This concludes our analysis leading to the result that the matter
multiplet fields of $\NN=4$ supergravity do not give any
logarithmic correction to the entropy of a quarter BPS black holes.
In fact since the heat kernel vanishes for all $s$, the full one loop
contribution from the massless matter multiplet 
vanishes.\footnote{We do not rule out the possibility of a finite
left-over contribution due to different ultra-violet cut-off on different
terms imposed by string theory.}
Since we have not computed
the contribution due to the gravity multiplet fields, our analysis
does not produce the complete
logarithmic correction to the entropy. Nevertheless our result
has non-trivial prediction for the entropy. For this
recall that there is a whole class of $\NN=4$ supersymmetric
string theories with different number of matter multiplet 
fields\cite{9508144,9508154}.
In these theories 
the quadratic action of fluctuating fields around the attractor
geometry will have exactly the same structure as discussed
here except that the index $r$ now runs over a lower number
of values than 22. 
Since the quadratic action of the gravity multiplet fields is common
to all these theories, the one loop contribution from these
fields to the entropy will also be identical. The vanishing of
the contribution from the matter sector then implies that for all
the $\NN=4$ supersymmetric theories the one loop contributions to
the black hole entropy from the masless fields are identical.
In particular the logarithmic corrections to the entropy -- which
we have argued earlier come only from the one loop
contribution due to the massless fields -- must also be identical.
This is consistent with the microscopic result for the entropy of 
quarter BPS states in a variety of $\NN=4$ supersymmetric
string theories. None of these theories have any logarithmic
correction to the entropy of quarter BPS black holes
irrespective of the number of matter multiplets they 
have\cite{0412287,0510147,0609109}.

\sectiono{Discussion} \label{sdis}

In this paper we have analyzed the eigenvalues and eigenfunctions
associated with the fluctuations of massless matter multiplet fields in the
near horizon geometry of quarter BPS black holes in $\NN=4$
supersymmetric string theories. This allows us to calculate the one
loop effective action and the logarithmic correction to the
Bekenstein-Hawking entropy due to the fields in the matter
multiplet. We find that even though individual fields contribute
to the effective action and logarithmic correction to the
black hole entropy, the net contribution from all the fields in
the matter multiplet vanishes. This is consistent with the fact
that there are no logarithmic corrections to the microscopic entropy
in $\NN=4$ supersymmetric string theories. In particular since the
logarithmic
contribution to the microscopic entropy vanishes 
independent of how many matter multiplet fields we have in
the theory, we
would have
run into an inconsistency if there had been a non-vanishing
logarithmic
contribution from the matter multiplet fields to the macroscopic
entropy.

Ref.\cite{duffroc} presented
a general analysis, based on the computation of the trace anomaly,
which showed that the trace anomaly vanishes for on-shell backgrounds
in $\NN=4$ and $\NN=8$ gauged supergravity theories. Since the trace
anomaly is related to the coefficient of the $s$ independent term in
the expansion of $K(0;s)$  via relations of the type described
in \refb{elocal1},  our result may appear to be similar in spirit
to those in \cite{duffroc}. However the analysis of
\cite{duffroc}, being a local analysis, does not take into account
the possible subtraction term given in \refb{edefkp} for removing the zero
modes. Indeed, the results of \cite{duffroc} would change if we
had replaced some of the fields by their dual description, {\it e.g.}
the scalars by 2-form fields. Our analysis shows the vanishing of
$K(0;s)$ for all $s$ and hence also the regulated $\wh K(0;s)$ given in
\refb{edefkp}. Since $\wh K(0;s)$  remains unchanged
when we replace a field by its dual description, 
the vanishing of $\wh K(0;s)$
holds irrespective of the duality frame we use to describe the fields.

One might naively conclude that the cancelation 
we have found is a result
of supersymmetry. However examining the microscopic results for the
black hole entropy we find that while quarter BPS black holes in
$\NN=4$ supersymmetric string theories have no logarithmic corrections
to their entropy, 1/8 BPS black holes in $\NN=8$ supersymmetric 
string theories, having the same amount of supersymmetry as the
quarter BPS black holes in $\NN=4$ supersymmetric string theories,
do have logarithmic corrections to their entropy. Thus the 
cancelation observed above cannot merely be a
consequence of supersymmetry. Nevertheless the vanishing of the
matter multiplet contribution to the logarithmic corrections
is crucial for correct matching with the microscopic entropy of quarter
BPS black holes in $\NN=4$ supersymetric string theories,
which do not have any logarithmic terms which depend on the
number of matter multiplets.

It is clearly desirable to extend the computation to include the 
fields in the gravity multiplet, both in the $\NN=4$ and $\NN=8$
supersymmetric string theories, and verify that the macroscopic results
are in agreement with the microscopic results given in
\refb{esmicro}. This can be done either
by the brute force approach of diagonalizing the fluctuations in the
gravity multiplet fields in the near horizon geometry, or possibly by
carrying out a direct string one loop calculation as in \cite{polc}.
The latter computation will give the complete one loop contribution,
including the order one contribution 
from the massive states, in one step. 
In this case the answer would be given by
an integration over the modular parameter $\tau$ of the torus, with
its imaginary part playing the role of the integration variable $s$
and the integrand a generalized version of $K(0;s)$ that also takes
into account the contribution from the massive modes.
The difficulty in carrying out this program lies in the fact
that we have to solve string theory in Ramond-Ramond
background and then carry out an exact one loop calculation in this
background. While this is not an easy task, it will be interesting to
see if the pure spinor formalism\cite{0910.2254}
or the hybrid formalism of \cite{9907200,0811.1758} can be of help.
An attempt to do this path integral using localization principle and
semi-classical method can be found in \cite{0608021}.

\bigskip

\noindent{\bf Acknowledgement:} We would like to thank Nabamita
Banerjee, Atish Dabholkar, Justin David,
Joao Gomes, Rajesh Gopakumar, Sameer Murthy 
and Boris Pioline for
useful discussions. 
The work of A.S. was supported in
part by the J. C.  Bose fellowship of the Department of
Science and Technology, India and by
the Chaires Internationales de Recherche Blaise Pascal, France.
S.B. would like to acknowledge IIT, Chennai
and Center For High Energy Physics, IISc Bangalore , where some
computations related to this work was performed.

\appendix

\sectiono{Analysis of the zero mode contribution} \label{sa}

In the analysis of \S\ref{sother} we had removed
the zero mode contribution while evaluating the determinant of the
kinetic operator for various fields. In this section we shall analyze the
result of the zero mode integrals for the vector fields on $AdS_2\times S^2$
-- the only fields containing zero modes which appeared in the
explicit analysis of \S\ref{sfive} -- and show that their contribution cancels
against another contribution that was left out in the analysis of
\S\ref{sfive}.

Let $A_\mu$ be a vector field of $AdS_2\times S^2$ and $g_{\mu\nu}$
be the background metric. This has the form
\be \label{eap1}
g_{\mu\nu} = a^2 \, g^{(0)}_{\mu\nu}\, ,
\ee
where $a$ is the size parameter of $S^2$ and $AdS_2$ 
and $g^{(0)}_{\mu\nu}$
is independent of $a$. Now in our analysis in
\S\ref{sother} we have assumed that the integration 
over $A_\mu$ gives the determinant of the kinetic operator
$(d\delta +\delta d)$ constructed from the metric $g_{\mu\nu}$. For this
we need to normalize the path integral over $A_\mu$ such that
\be \label{eap2}
\int [DA_\mu] \exp\left[- \int d^4 x \, \sqrt{\det g} \, g^{\mu\nu} A_\mu A_\nu
\right] = 1\, .
\ee
Using \refb{eap1} this may be expressed as
\be \label{eap3}
\int [DA_\mu] \exp\left[- a^2 \int d^4 x \, \sqrt{\det g^{(0)}} \, 
g^{(0)\mu\nu} A_\mu A_\nu
\right] = 1\, .
\ee
{}From this we see that up to an $a$ independent normalization
constant, $[DA_\mu]$  actually corresponds to integration
with measure $\prod_x d(a A_\mu(x))$. This in turn implies that
integration over every zero mode of $A_\mu(x)$ with the measure
induced from $[DA_\mu]$ will produce a factor
of $a$.

Now for a non-zero mode, the path integral weighted by the exponential
of the action produces a factor of $\kappa_n^{-1/2}$ where $\kappa_n$
is the eigenvalue of the kinetic operator. Since $\kappa_n$ has the form
$b_n/a^2$ where $b_n$ is an $a$ independent constant, integration
over a non-zero mode produces a factor proportional to $a$. Thus when
we remove the contribution due to the zero modes, we remove a factor
of $a$ for each zero mode. However the analysis of the previous paragraph
showed that integration over the zero modes gives us back a factor of $a$.
Thus the net result is that for computing the coefficient
of the $\ln a$ term we can effectively ignore the
subtraction described in \refb{edefkp}
and continue to use the full heat kernel $K^v(0;s)$ {\it provided we
use the prescription that the $\ln a$ terms arise from
integration over $s$ in the range $1<< s << a^2$ even though the
integral $\int{ds\, s^{-1}} K^v(0;s)$
does not converge at large  $s$.}

Looking back at our final expression \refb{enetv} for the net contribution
to the trace of the heat kernel from a matter multiplet, we see that the
right hand side of this expression in fact vanishes rapidly as $s\to\infty$
since it is proportional to the scalar heat kernel in $AdS_2$ which does
not have any zero mode contribution. So indeed we did not need to explicitly
carry out any subtraction of the type given in \refb{edefkp}. Technically this
was due to the fact that there was another term that approached a
constant as $s\to\infty$ and canceled the constant term in the trace of
the vector heat kernel in $AdS_2\times S^2$. This new term arose from
the product of $K^s_{S^2}$ and $\delta K^{v+s}_{AdS_2}$ given in
\refb{eshift7a}. So if we can argue that the correct prescription for
evaluating the contribution from $K^s_{S^2} \delta K^{v+s}_{AdS_2}$
to the $\ln a$ terms is to {\it not subtract the constant term as $s\to\infty$},
and restrict the integration over $s$ to the range $1<<s << a^2$,
then our final result \refb{enetv} will be justified; we do not subtract 
any constant either from the original $K^v_{AdS_2\times S^2}$, nor
from the correction term $K^s_{S^2} \delta K^{v+s}_{AdS_2}$.

Thus our task now is to justify \refb{eshift7a} for calculating the effect
of the flux in $AdS_2$ without any subtraction. 
If we adopt this prescription then the net
change proportional to
$\ln a$ in $-{1\over 2}
\ln \det (d\delta +\delta d)$ 
due to the presence of the flux through $AdS_2$ will be
given by
\be \label{eap4pre}
 {1\over 2} \, \int d^4 x \sqrt{\det g}\,
 \int_{1<< s<< a^2} \, {ds\over s}\, 
K^s_{S^2}(0;s)\delta K^{v+s}_{AdS_2} (0;s)\, .
\ee
If we denote by $u$ and $v$ the coordinates of $AdS_2$ and
$S^2$ respectively, and
pick a particular eigenfunction on $S^2$ with eigenvalue $c/a^2$
and eigenfunction $f(v)$, then the contribution from this eigenfunction
on $S^2$ to \refb{eap4pre} will be given by
\be \label{eap4}
{1\over 2} \, \int d^4 x \, |f(v)|^2 \sqrt{\det g}\,
 \int_{1<< s<< a^2} \, {ds\over s} \, e^{-cs/a^2}\, 
 \delta K^{v+s}_{AdS_2} (0;s)
= -{1\over 4\pi a^2} \, \ln (a^2)\, \int d^4 x 
\, |f(v)|^2\, \sqrt{\det g}\, .
\ee
Note that we have used the ad hoc prescription of
restricting the integration range to $1<< s<< a^2$;
without this the integral will diverge from the large $s$
region for $c=0$.
Now we shall verify the correctness of this result using an
independent procedure that does not require this
ad hoc prescription. For this we
go back to \refb{eshift4}. From this equation it is clear that the effect of the
flux is to take a pair of eigenvalues 
$\left(c+\lambda^2+{1\over 4}\right)/a^2$
of $-\square_{S^2}-\square_{AdS_2}$  and shift them to
$\left(c+(\lambda\pm i)^2 +{1\over 4}\right)/a^2$. 
Now since we are interested in computing
the determinant, we could also interpret this as shifting a factor of
$\left(c+\lambda^2+{1\over 4}\right)^2/a^4$ in the determinant to
$\left|c+(\lambda+ i)^2 +{1\over 4}\right|^2/a^4$. Thus the change in
$-{1\over 2} \ln\det (d\delta +\delta d)$ can be written as
\ben \label{eap5}
&& {1\over 4\pi a^2}\, \int_{\tilde\eps}^\infty 
{dt\over t} \, \int d^4 x \, |f(v)|^2 \sqrt{\det g} \, 
\, \int_0^\infty \, d\lambda \, \lambda \, 
\tanh(\pi \lambda) \nonumber \\
&& \quad \left[ \exp\left( - t \left|c+(\lambda+ i)^2 +{1\over 4}\right|^2/a^4\right)
- \exp\left( - t \left(c+\lambda^2+{1\over 4}\right)^2/a^4 \right)
\right]\, ,
\een
where we have used the fact that the distribution function of the parameter
$\lambda$ is given by $\lambda \tanh(\pi\lambda) / (2\pi a^2)$.
$\tilde\eps$ is an untraviolet cut-off of order 1.
This integral is manifestly convergent at large $t$ 
even for $c=0$ and does not have the
problem mentioned in footnote \ref{ff}.
Since the integrand falls off rapidly for $t>>a^4$, the possible $\ln a$
terms come from integration over the range $1<<t<<a^4$.
Using the method described in \S\ref{s2} one can estimate the behaviour
of the integrand in this range after carrying out the
$\lambda$ integral, and finds the result:
\be \label{eap6}
-{1\over 8\pi a^2} \, \int d^4 x \, |f(v)|^2 \sqrt{\det g}
\, \int_{1<< t<< a^{4}}
{dt\over t} \simeq 
 -{1\over 4\pi a^2} \, \ln (a^2)\, \int d^4 x \, |f(v)|^2 \sqrt{\det g}\, .
\ee
This is in perfect agreement with \refb{eap4}, showing that the
prescription of using the full result $\delta K^{v+s}_{AdS_2}$ given in
\refb{eshift7a} without any subtraction and restricting the integration
in the range $1<<s<<a^2$ gives the correct $\ln a$ factors in the
determinant. Of course since 
the final result \refb{enetv} falls off sufficient
rapidly for $s\to\infty$ we can drop the requirement of restricting the
integration to the range $s<<a^2$.

This concludes our proof that even after taking into account the possible
additional factors of $\ln a$ which could arise from zero mode integration,
\refb{enetv} can be used to calculate the logarithmic correction to the
black hole entropy due to the bosonic fields in the matter multiplet.


\begin{thebibliography}{99} 

\baselineskip=12pt 
\parskip=0pt

\small 

 \bibitem{9307038}
  R.~M.~Wald,
  ``Black hole entropy in the Noether charge,''
  Phys.\ Rev.\ D {\bf 48}, 3427 (1993)
  [arXiv:gr-qc/9307038].

\bibitem{9312023}
  T.~Jacobson, G.~Kang and R.~C.~Myers,
  ``On Black Hole Entropy,''
  Phys.\ Rev.\ D {\bf 49}, 6587 (1994)
  [arXiv:gr-qc/9312023].

\bibitem{9403028}
  V.~Iyer and R.~M.~Wald,
  ``Some properties of Noether 
  charge and a proposal for dynamical black hole
  entropy,''
  Phys.\ Rev.\ D {\bf 50}, 846 (1994)
  [arXiv:gr-qc/9403028].

\bibitem{9502009}
  T.~Jacobson, G.~Kang and R.~C.~Myers,
  ``Black hole entropy in higher curvature gravity,''
  arXiv:gr-qc/9502009.


\bibitem{0506177}
  A.~Sen,
  ``Black hole entropy function and the attractor mechanism in higher
  derivative gravity,''
  JHEP {\bf 0509}, 038 (2005)
  [arXiv:hep-th/0506177].

\bibitem{0508042}
  A.~Sen,
  ``Entropy function for heterotic black holes,''
  JHEP {\bf 0603}, 008 (2006)
  [arXiv:hep-th/0508042].
  
\bibitem{9508072}
  S.~Ferrara, R.~Kallosh and A.~Strominger,
  ``N=2 extremal black holes,''
  Phys.\ Rev.\ D {\bf 52}, 5412 (1995)
  [arXiv:hep-th/9508072].

\bibitem{9602111}
  A.~Strominger,
  ``Macroscopic Entropy of $N=2$ Extremal Black Holes,''
  Phys.\ Lett.\ B {\bf 383}, 39 (1996)
  [arXiv:hep-th/9602111].

\bibitem{9602136}
  S.~Ferrara and R.~Kallosh,
  ``Supersymmetry and Attractors,''
  Phys.\ Rev.\ D {\bf 54}, 1514 (1996)
  [arXiv:hep-th/9602136].

\bibitem{0903.1477}
  A.~Sen,
  ``Arithmetic of Quantum Entropy Function,''
  JHEP {\bf 0908}, 068 (2009)
  [arXiv:0903.1477 [hep-th]].
  
\bibitem{appear}
A.~Dabholkar, J.~Gomes, S.~Murthy and A.~Sen, to appear.

\bibitem{0412287}
  G.~Lopes Cardoso, B.~de Wit, J.~Kappeli and T.~Mohaupt,
  ``Asymptotic degeneracy of dyonic N = 4 string states and black hole
  JHEP {\bf 0412}, 075 (2004)
  [arXiv:hep-th/0412287].

\bibitem{0510147}
  D.~P.~Jatkar and A.~Sen,
  ``Dyon spectrum in CHL models,''
  JHEP {\bf 0604}, 018 (2006)
  [arXiv:hep-th/0510147].

\bibitem{0605210}
  J.~R.~David and A.~Sen,
  ``CHL dyons and statistical entropy function from D1-D5 system,''
  arXiv:hep-th/0605210.

\bibitem{0609109}
  J.~R.~David, D.~P.~Jatkar and A.~Sen,
  ``Dyon spectrum in generic N = 4 supersymmetric Z(N) orbifolds,''
  arXiv:hep-th/0609109.

\bibitem{9607026}
R.~Dijkgraaf, E.~P.~Verlinde and H.~L.~Verlinde,
``Counting dyons in N = 4 string theory,''
Nucl.\ Phys.\ B {\bf 484}, 543 (1997)
[arXiv:hep-th/9607026].

\bibitem{0505094}
D.~Shih, A.~Strominger and X.~Yin,
``Recounting dyons in N = 4 string theory,''
arXiv:hep-th/0505094.

\bibitem{0506249}
D.~Gaiotto,
``Re-recounting dyons in N = 4 string theory,''
arXiv:hep-th/0506249.

\bibitem{0508174}
  D.~Shih and X.~Yin,
  ``Exact black hole degeneracies and the topological string,''
  JHEP {\bf 0604}, 034 (2006)
  [arXiv:hep-th/0508174].

\bibitem{0602254}
  J.~R.~David, D.~P.~Jatkar and A.~Sen,
  ``Product representation of dyon partition function in CHL models,''
  JHEP {\bf 0606}, 064 (2006)
  [arXiv:hep-th/0602254].

\bibitem{0603066}
  A.~Dabholkar and S.~Nampuri,
  ``Spectrum of dyons and black holes in
  CHL orbifolds using Borcherds lift,''
  arXiv:hep-th/0603066.

\bibitem{0607155}
  J.~R.~David, D.~P.~Jatkar and A.~Sen,
  ``Dyon spectrum in N = 4 supersymmetric type II string theories,''
  arXiv:hep-th/0607155.

\bibitem{0612011}
  A.~Dabholkar and D.~Gaiotto,
  ``Spectrum of CHL dyons from genus-two partition function,''
  arXiv:hep-th/0612011.

\bibitem{0702141}
  A.~Sen,
  ``Walls of marginal stability and dyon spectrum in N = 4 supersymmetric
  string theories,''
  arXiv:hep-th/0702141.

\bibitem{0702150}
  A.~Dabholkar, D.~Gaiotto and S.~Nampuri,
  ``Comments on the spectrum of CHL dyons,''
  arXiv:hep-th/0702150.

\bibitem{0705.1433}
  N.~Banerjee, D.~P.~Jatkar and A.~Sen,
  ``Adding charges to N = 4 dyons,''
  arXiv:0705.1433 [hep-th].

\bibitem{0705.3874}
  A.~Sen,
  ``Two Centered Black Holes and N=4 Dyon Spectrum,''
  arXiv:0705.3874 [hep-th].

\bibitem{0706.2363}
  M.~C.~N.~Cheng and E.~Verlinde,
  ``Dying Dyons Don't Count,''
  arXiv:0706.2363 [hep-th].

\bibitem{0708.1270}
  A.~Sen,
  ``Black Hole Entropy Function,
Attractors and Precision Counting of
  Microstates,''
Gen.\ Rel.\ Grav.\  {\bf 40}, 2249 (2008)
  [arXiv:0708.1270 [hep-th]].


\bibitem{0802.0544}
  S.~Banerjee, A.~Sen and Y.~K.~Srivastava,
  ``Generalities of Quarter BPS Dyon
Partition Function and Dyons of Torsion
  Two,''
  arXiv:0802.0544 [hep-th].

\bibitem{0802.1556}
  S.~Banerjee, A.~Sen and Y.~K.~Srivastava,
  ``Partition Functions of Torsion $>1$ Dyons in Heterotic
String Theory on $T^6$,''
  arXiv:0802.1556 [hep-th].

\bibitem{0803.2692}
  A.~Dabholkar, J.~Gomes and S.~Murthy,
  ``Counting all dyons in N =4 string theory,''
  arXiv:0803.2692 [hep-th].

\bibitem{0806.2337}
  M.~C.~N.~Cheng and E.~P.~Verlinde,
  ``Wall Crossing, Discrete Attractor Flow, and Borcherds Algebra,''
  arXiv:0806.2337 [hep-th].

\bibitem{0807.4451}
  S.~Govindarajan and K.~Gopala Krishna,
  ``Generalized Kac-Moody Algebras from CHL dyons,''
  JHEP {\bf 0904}, 032 (2009)
  [arXiv:0807.4451 [hep-th]].

\bibitem{0809.4258}
  M.~C.~N.~Cheng and A.~Dabholkar,
  ``Borcherds-Kac-Moody Symmetry of N=4 Dyons,''
  arXiv:0809.4258 [hep-th].

\bibitem{0808.1746}
  S.~Banerjee, A.~Sen and Y.~K.~Srivastava,
  ``Genus Two Surface and Quarter BPS Dyons: The Contour Prescription,''
  JHEP {\bf 0903}, 151 (2009)
  [arXiv:0808.1746 [hep-th]].

\bibitem{0901.1758}
  M.~C.~N.~Cheng and L.~Hollands,
  ``A Geometric Derivation of the Dyon Wall-Crossing Group,''
  JHEP {\bf 0904}, 067 (2009)
  [arXiv:0901.1758 [hep-th]].

\bibitem{0907.1410}
  S.~Govindarajan and K.~Gopala Krishna,
  ``BKM Lie superalgebras from dyon spectra in $Z_N$
  CHL orbifolds for composite
  N,''
  arXiv:0907.1410 [hep-th].

\bibitem{0911.0586}
  A.~Dabholkar and J.~Gomes,
  ``Perturbative tests of non-perturbative counting,''
  arXiv:0911.0586 [hep-th].

\bibitem{0911.1563}
  A.~Sen,
  ``A Twist in the Dyon Partition Function,''
  arXiv:0911.1563 [hep-th].

\bibitem{1002.3857}
  A.~Sen,
  ``Discrete Information from CHL Black Holes,''
  arXiv:1002.3857 [hep-th].
  
  
\bibitem{suresh}
S.~Govindarajan, 
"BKM Lie superalgebras from twisted
CHL dyons", to appear.

\bibitem{9507090}
  M.~Cvetic and D.~Youm,
  ``Dyonic BPS saturated black holes of heterotic string on a six torus,''
  Phys.\ Rev.\  D {\bf 53}, 584 (1996)
  [arXiv:hep-th/9507090].

\bibitem{9512031}
M.~Cvetic and A.~A.~Tseytlin,
``Solitonic strings and BPS saturated dyonic black holes,''
  Phys.\ Rev.\  D {\bf 53}, 5619 (1996)
  [Erratum-ibid.\  D {\bf 55}, 3907 (1997)]
  [arXiv:hep-th/9512031].

\bibitem{9711053}
  J.~M.~Maldacena, A.~Strominger and E.~Witten,
  ``Black hole entropy in M-theory,''
  JHEP {\bf 9712}, 002 (1997)
  [arXiv:hep-th/9711053].

\bibitem{9812082}
G.~Lopes Cardoso, B.~de Wit and T.~Mohaupt,
``Corrections to macroscopic supersymmetric black-hole entropy,''
Phys.\ Lett.\ B {\bf 451}, 309 (1999)
[arXiv:hep-th/9812082].

\bibitem{9906094}
  G.~Lopes Cardoso, B.~de Wit and T.~Mohaupt,
  ``Macroscopic entropy formulae and 
  non-holomorphic corrections for
  supersymmetric black holes,''
  Nucl.\ Phys.\  B {\bf 567}, 87 (2000)
  [arXiv:hep-th/9906094].

\bibitem{0007195}
T.~Mohaupt,
``Black hole entropy, special geometry and strings,''
Fortsch.\ Phys.\  {\bf 49}, 3 (2001)
[arXiv:hep-th/0007195].



\bibitem{1003.1083}
  R.~Aros, D.~E.~Diaz and A.~Montecinos,
  ``Logarithmic correction to BH entropy as Noether charge,''
  arXiv:1003.1083 [hep-th].

\bibitem{0809.3304}
  A.~Sen,
  ``Quantum Entropy Function from AdS(2)/CFT(1) Correspondence,''
  Int.\ J.\ Mod.\ Phys.\  A {\bf 24}, 4225 (2009)
  [arXiv:0809.3304 [hep-th]].

\bibitem{9803002}
  J.~M.~Maldacena,
  ``Wilson loops in large N field theories,''
  Phys.\ Rev.\ Lett.\  {\bf 80}, 4859 (1998)
  [arXiv:hep-th/9803002].

\bibitem{9803001}
  S.~J.~Rey and J.~T.~Yee,
  ``Macroscopic strings as heavy quarks 
  in large N gauge theory and  anti-de
  Sitter supergravity,''
  Eur.\ Phys.\ J.\  C {\bf 22}, 379 (2001)
  [arXiv:hep-th/9803001].

\bibitem{0904.4486}
  J.~Gomis, T.~Okuda and D.~Trancanelli,
  ``Quantum 't Hooft operators and 
  S-duality in N=4 super Yang-Mills,''
  arXiv:0904.4486 [hep-th].

\bibitem{0810.3472}
  N.~Banerjee, D.~P.~Jatkar and A.~Sen,
  ``Asymptotic Expansion of the N=4 Dyon Degeneracy,''
  JHEP {\bf 0905}, 121 (2009)
  [arXiv:0810.3472 [hep-th]].

\bibitem{0904.4253}
  S.~Murthy and B.~Pioline,
  ``A Farey tale for N=4 dyons,''
  JHEP {\bf 0909}, 022 (2009)
  [arXiv:0904.4253 [hep-th]].
 
\bibitem{0908.0039}
  A.~Sen,
  ``Arithmetic of N=8 Black Holes,''
  JHEP {\bf 1002}, 090 (2010)
  [arXiv:0908.0039 [hep-th]].

\bibitem{heckman}
  J.~J.~Duistermaat and G.~J.~Heckman,
  ``On The Variation In The 
  Cohomology Of The Symplectic Form Of The Reduced
  Phase Space,''
  Invent.\ Math.\  {\bf 69}, 259 (1982).

\bibitem{wittens}
  E.~Witten,
  ``Topological Quantum Field Theory,''
  Commun.\ Math.\ Phys.\  {\bf 117}, 353 (1988).

\bibitem{witten92}
  E.~Witten,
  ``The N Matrix Model And Gauged WZW Models,''
  Nucl.\ Phys.\  B {\bf 371}, 191 (1992).

\bibitem{9112056}
  E.~Witten,
  ``Mirror manifolds and topological field theory,''
  arXiv:hep-th/9112056.

\bibitem{9204083}
  E.~Witten,
  ``Two-dimensional gauge theories revisited,''
  J.\ Geom.\ Phys.\  {\bf 9} (1992) 303
  [arXiv:hep-th/9204083].

\bibitem{9511112}
  A.~S.~Schwarz and O.~Zaboronsky,
  ``Supersymmetry and localization,''
  Commun.\ Math.\ Phys.\  {\bf 183}, 463 (1997)
  [arXiv:hep-th/9511112].

\bibitem{zaboronsky}
O.~Zaboronsky,
``Dimensional reduction in supersymmetric field theories,''
J.\ Phys.\ {\bf A35}, 5511 (2002).

\bibitem{0206161}
  N.~A.~Nekrasov,
  ``Seiberg-Witten Prepotential From Instanton Counting,''
  Adv.\ Theor.\ Math.\ Phys.\  {\bf 7}, 831 (2004)
  [arXiv:hep-th/0206161].

\bibitem{0712.2824}
  V.~Pestun,
  ``Localization of gauge theory on a 
  four-sphere and supersymmetric Wilson
  loops,''
  arXiv:0712.2824 [hep-th].
 
 \bibitem{ati1}
M.F.~Atiyah, Elliptic operators and compact groups. Springer-Verlag,
Berlin,
1974. 

\bibitem{ati2}
P.~Shanahan, The atiyah-singer index theorem : An
introduction, Springer-Verlag.

\bibitem{0608021}
  C.~Beasley, D.~Gaiotto, M.~Guica, L.~Huang, 
  A.~Strominger and X.~Yin,
  ``Why Z(BH) = |Z(top)|**2,''
  arXiv:hep-th/0608021.

\bibitem{0905.2686}
  N.~Banerjee, S.~Banerjee, R.~Gupta, I.~Mandal and A.~Sen,
  ``Supersymmetry, Localization and Quantum Entropy Function,''
  arXiv:0905.2686 [hep-th].

\bibitem{0002040}
  R.~K.~Kaul and P.~Majumdar,
  ``Logarithmic correction to the Bekenstein-Hawking entropy,''
  Phys.\ Rev.\ Lett.\  {\bf 84}, 5255 (2000)
  [arXiv:gr-qc/0002040].

\bibitem{0005017}
  S.~Carlip,
  ``Logarithmic corrections to black hole entropy from the Cardy formula,''
  Class.\ Quant.\ Grav.\  {\bf 17}, 4175 (2000)
  [arXiv:gr-qc/0005017].

\bibitem{0104010}
  T.~R.~Govindarajan, R.~K.~Kaul and V.~Suneeta,
  ``Logarithmic correction to the 
  Bekenstein-Hawking entropy of the BTZ  black
  hole,''
  Class.\ Quant.\ Grav.\  {\bf 18}, 2877 (2001)
  [arXiv:gr-qc/0104010].

\bibitem{9708062}
  A.~Gregori, E.~Kiritsis, C.~Kounnas, N.~A.~Obers, 
  P.~M.~Petropoulos and B.~Pioline,
  ``R**2 corrections and non-perturbative 
  dualities of N = 4 string ground
  states,''
  Nucl.\ Phys.\ B {\bf 510}, 423 (1998)
  [arXiv:hep-th/9708062].

\bibitem{9708130}
  E.~Kiritsis,
  ``Introduction to non-perturbative string theory,''
  arXiv:hep-th/9708130.


\bibitem{0804.1773}
  S.~Giombi, A.~Maloney and X.~Yin,
  ``One-loop Partition Functions of 3D Gravity,''
  JHEP {\bf 0808}, 007 (2008)
  [arXiv:0804.1773 [hep-th]].
  
\bibitem{0911.5085}
  J.~R.~David, M.~R.~Gaberdiel and R.~Gopakumar,
  ``The Heat Kernel on $AdS_3$ and its Applications,''
  arXiv:0911.5085 [hep-th].

\bibitem{9709064}
  R.~B.~Mann and S.~N.~Solodukhin,
  ``Universality of quantum entropy for extreme black holes,''
  Nucl.\ Phys.\  B {\bf 523}, 293 (1998)
  [arXiv:hep-th/9709064].

\bibitem{9610237}
  J.~A.~Harvey and G.~W.~Moore,
  ``Fivebrane instantons and R**2 couplings in N = 4 string theory,''
  Phys.\ Rev.\  D {\bf 57}, 2323 (1998)
  [arXiv:hep-th/9610237].


\bibitem{9412161}
  D.~V.~Fursaev,
  ``Temperature And Entropy Of A 
  Quantum Black Hole And Conformal Anomaly,''
  Phys.\ Rev.\  D {\bf 51}, 5352 (1995)
  [arXiv:hep-th/9412161].
  
\bibitem{9604118}
  R.~B.~Mann and S.~N.~Solodukhin,
  ``Conical geometry and quantum entropy of a charged Kerr black hole,''
  Phys.\ Rev.\  D {\bf 54}, 3932 (1996)
  [arXiv:hep-th/9604118].

\bibitem{0406044}
  A.~J.~M.~Medved,
  ``A comment on black hole entropy or why Nature abhors a logarithm,''
  Class.\ Quant.\ Grav.\  {\bf 22}, 133 (2005)
  [arXiv:gr-qc/0406044].

\bibitem{0409024}
  D.~N.~Page,
  ``Hawking radiation and black hole thermodynamics,''
  New J.\ Phys.\  {\bf 7}, 203 (2005)
  [arXiv:hep-th/0409024].


\bibitem{0805.2220}
  R.~Banerjee and B.~R.~Majhi,
  ``Quantum Tunneling Beyond Semiclassical Approximation,''
  JHEP {\bf 0806}, 095 (2008)
  [arXiv:0805.2220 [hep-th]].

\bibitem{0808.3688}
  R.~Banerjee and B.~R.~Majhi,
  ``Quantum Tunneling, Trace Anomaly and Effective Metric,''
  Phys.\ Lett.\  B {\bf 674}, 218 (2009)
  [arXiv:0808.3688 [hep-th]].

\bibitem{0809.1508}
  B.~R.~Majhi,
  ``Fermion Tunneling Beyond Semiclassical Approximation,''
  Phys.\ Rev.\  D {\bf 79}, 044005 (2009)
  [arXiv:0809.1508 [hep-th]].

\bibitem{0911.4379}
  R.~G.~Cai, L.~M.~Cao and N.~Ohta,
  ``Black Holes in Gravity with 
  Conformal Anomaly and Logarithmic Term in Black
  Hole Entropy,''
  arXiv:0911.4379 [hep-th].
  
\bibitem{gilkey}
P.~B.~Gilkey, 
``Invariance theory, the heat equation and the Atiyah-Singer index theorem,''
Publish or Perish Inc., USA (1984).


\bibitem{0306138}
  D.~V.~Vassilevich,
  ``Heat kernel expansion: User's manual,''
  Phys.\ Rept.\  {\bf 388}, 279 (2003)
  [arXiv:hep-th/0306138].

\bibitem{0305010}
  K.~Kirsten and A.~J.~McKane,
  ``Functional determinants by contour integration methods,''
  Annals Phys.\  {\bf 308}, 502 (2003)
  [arXiv:math-ph/0305010].

\bibitem{0602106}
  T.~Hartman and L.~Rastelli,
  ``Double-trace deformations, mixed boundary conditions and functional
  determinants in AdS/CFT,''
  JHEP {\bf 0801}, 019 (2008)
  [arXiv:hep-th/0602106].

\bibitem{0908.2657}
  F.~Denef, S.~A.~Hartnoll and S.~Sachdev,
  ``Black hole determinants and quasinormal modes,''
  arXiv:0908.2657 [hep-th].


\bibitem{campo}
  R.~Camporesi,
  ``Harmonic analysis and propagators on homogeneous spaces,''
  Phys.\ Rept.\  {\bf 196} (1990) 1.

\bibitem{camhig1}
  R.~Camporesi and A.~Higuchi,
  ``Spectral functions and zeta functions in hyperbolic spaces,''
  J.\ Math.\ Phys.\  {\bf 35}, 4217 (1994).

\bibitem{campo2}
  R.~Camporesi,
  ``The Spinor heat kernel in maximally symmetric spaces,''
  Commun.\ Math.\ Phys.\  {\bf 148} (1992) 283.

\bibitem{camhig2}
  R.~Camporesi and A.~Higuchi,
  ``Arbitrary spin effective potentials in anti-de Sitter space-time,''
  Phys.\ Rev.\  D {\bf 47}, 3339 (1993).

\bibitem{9408068}
  S.~N.~Solodukhin,
  ``On 'Nongeometric' contribution 
  to the entropy of black hole due to quantum
  corrections,''
  Phys.\ Rev.\  D {\bf 51}, 618 (1995)
  [arXiv:hep-th/9408068].

\bibitem{9407001}
  S.~N.~Solodukhin,
  ``The Conical singularity and quantum corrections to entropy of black hole,''
  Phys.\ Rev.\  D {\bf 51}, 609 (1995)
  [arXiv:hep-th/9407001].

\bibitem{dewitt}
B. S. DeWitt, Dynamical Theory of Groups and Fields, 
Gordon and Breach,
New York, 1965.

\bibitem{mckean}
H. P. McKean and I. M. Singer, ÒCurvature and the eigenvalues of the
Laplacian,Ó J. Diff. Geom. 1 (1967) 43Ð69.

\bibitem{christ-duff1}
  S.~M.~Christensen and M.~J.~Duff,
  ``New Gravitational Index Theorems And Supertheorems,''
  Nucl.\ Phys.\  B {\bf 154}, 301 (1979).

\bibitem{christ-duff2}
  S.~M.~Christensen and M.~J.~Duff,
  ``Quantizing Gravity With A Cosmological Constant,''
  Nucl.\ Phys.\  B {\bf 170}, 480 (1980).

\bibitem{0908.3402}
  A.~Sen,
  ``Two Charge System Revisited: Small 
  Black Holes or Horizonless Solutions?,''
  arXiv:0908.3402 [hep-th].

\bibitem{9505009}
  R.~Camporesi and A.~Higuchi,
  ``On The Eigen Functions Of 
  The Dirac Operator On Spheres And Real Hyperbolic
  Spaces,''
  J.\ Geom.\ Phys.\  {\bf 20}, 1 (1996)
  [arXiv:gr-qc/9505009].


\bibitem{9505186}
  E.~Witten,
  ``On S duality in Abelian gauge theory,''
  Selecta Math.\  {\bf 1} (1995) 383
  [arXiv:hep-th/].

\bibitem{duffnieu}
  M.~J.~Duff and P.~van Nieuwenhuizen,
  ``Quantum Inequivalence Of Different Field Representations,''
  Phys.\ Lett.\  B {\bf 94}, 179 (1980).


\bibitem{siegel}
  W.~Siegel,
  ``Hidden Ghosts,''
  Phys.\ Lett.\  B {\bf 93}, 170 (1980).

\bibitem{ThierryMieg}
  J.~Thierry-Mieg,
  ``Brs Structure Of The Antisymmetric Tensor Gauge Theories,''
  Nucl.\ Phys.\  B {\bf 335}, 334 (1990).

\bibitem{sezgin}
  E.~Sezgin and P.~van Nieuwenhuizen,
  ``Renormalizability Properties Of 
  Antisymmetric Tensor Fields Coupled To
  Gravity,''
  Phys.\ Rev.\  D {\bf 22}, 301 (1980).

\bibitem{0806.3505}
  I.~L.~Buchbinder, E.~N.~Kirillova and N.~G.~Pletnev,
  ``Quantum Equivalence of 
  Massive Antisymmetric Tensor Field Models in Curved
  Space,''
  Phys.\ Rev.\  D {\bf 78}, 084024 (2008)
  [arXiv:0806.3505 [hep-th]].

\bibitem{GSW}
  M.~B.~Green, J.~H.~Schwarz and E.~Witten,
  ``Superstring Theory. Vol. 2: Loop Amplitudes, Anomalies And Phenomenology,''
{\it  Cambridge, Uk: Univ. Pr. ( 1987) 596 P. ( Cambridge Monographs On Mathematical Physics)}


\bibitem{9207016}
  J.~Maharana and J.~H.~Schwarz,
  ``Noncompact symmetries in string theory,''
  Nucl.\ Phys.\  B {\bf 390} (1993) 3
  [arXiv:hep-th/9207016].

\bibitem{9402002}
  A.~Sen,
  ``Strong - weak coupling duality in four-dimensional string theory,''
  Int.\ J.\ Mod.\ Phys.\  A {\bf 9}, 3707 (1994)
  [arXiv:hep-th/9402002].

\bibitem{1001.2933}
  E.~Witten,
  ``Analytic Continuation Of Chern-Simons Theory,''
  arXiv:1001.2933 [hep-th].


\bibitem{9508144}
S.~Chaudhuri and D.~A.~Lowe,
``Type IIA heterotic duals with maximal supersymmetry,''
Nucl.\ Phys.\ B {\bf 459}, 113 (1996)
[arXiv:hep-th/9508144].

\bibitem{9508154}
P.~S.~Aspinwall,
``Some relationships between dualities in string theory,''
Nucl.\ Phys.\ Proc.\ Suppl.\  {\bf 46}, 30 (1996)
[arXiv:hep-th/9508154].

\bibitem{duffroc}
  S.~M.~Christensen, M.~J.~Duff, G.~W.~Gibbons and M.~Rocek,
  ``Vanishing One Loop Beta Function In Gauged N $>$ 4 Supergravity,''
  Phys.\ Rev.\ Lett.\  {\bf 45}, 161 (1980).

\bibitem{polc}
  J.~Polchinski,
  ``Evaluation Of The One Loop String Path Integral,''
  Commun.\ Math.\ Phys.\  {\bf 104}, 37 (1986).

\bibitem{0910.2254}
  O.~A.~Bedoya and N.~Berkovits,
  ``GGI Lectures on the Pure Spinor Formalism of the Superstring,''
  arXiv:0910.2254 [hep-th].

\bibitem{9907200}
  N.~Berkovits, M.~Bershadsky, T.~Hauer, S.~Zhukov and B.~Zwiebach,
  ``Superstring theory on AdS(2) x S(2) as a coset supermanifold,''
  Nucl.\ Phys.\  B {\bf 567}, 61 (2000)
  [arXiv:hep-th/9907200].

\bibitem{0811.1758}
  B.~Chandrasekhar,
  ``Black Hole Partition Function using Hybrid Formalism of Superstrings,''
  arXiv:0811.1758 [hep-th].


 \end{thebibliography}
\end{document}